\documentclass[manuscript,acmsmall,nonacm]{acmart}

\usepackage{xcolor}
\usepackage{enumitem}
\usepackage{balance}
\usepackage{caption}
\usepackage{subcaption}

\newcounter{guideline}
\newcommand{\guideline}{\refstepcounter{guideline}\textbf{(G\theguideline})}

\AtBeginDocument{%
  \providecommand\BibTeX{{%
    \normalfont B\kern-0.5em{\scshape i\kern-0.25em b}\kern-0.8em\TeX}}}

\setcopyright{acmcopyright}
\copyrightyear{2018}
\acmYear{2018}
\acmDOI{XXXXXXX.XXXXXXX}

\acmConference[CSCW 'XX]{Conf title}{June 03--05,
 2018}{Woodstock, NY}

\begin{document}

\title[Factory Operators' Perspectives on Cognitive Assistants]{Factory Operators' Perspectives on Cognitive Assistants for Knowledge Sharing: Challenges, Risks, and Impact on Work}

\author{S. Kernan Freire}
\orcid{0000-0001-8684-0585}
\affiliation{%
  \institution{Delft University of Technology}
  \streetaddress{Landbergstraat 15}
  \city{Delft}
  \postcode{2628 CE}
  \country{The Netherlands}}
\email{s.kernanfreire@tudelft.nl}

\author{T. He}
\orcid{0000-0002-8486-3940}
\affiliation{%
  \institution{Delft University of Technology}
  \streetaddress{Landbergstraat 15}
  \city{Delft}
  \postcode{2628 CE}
  \country{The Netherlands}}
\email{t.he-1@tudelft.nl}

\author{C. Wang}
\orcid{0000-0001-8213-6582}
\affiliation{%
  \institution{Delft University of Technology}
  \streetaddress{Landbergstraat 15}
  \city{Delft}
  \postcode{2628 CE}
  \country{The Netherlands}}
\email{c.wang-16@tudelft.nl}

\author{E. Niforatos}
\orcid{0000-0002-0484-4214}
\affiliation{%
  \institution{Delft University of Technology}
  \streetaddress{Landbergstraat 15}
  \city{Delft}
  \country{Netherlands}}
\email{e.niforatos@tudelft.nl}

\author{Alessandro Bozzon}
\orcid{0000-0002-3300-2913}
\affiliation{%
  \institution{Delft University of Technology}
  \streetaddress{Landbergstraat 15}
  \city{Delft}
  \country{Netherlands}}
\email{a.bozzon@tudelft.nl}

\renewcommand{\shortauthors}{Kernan Freire, et al.}

\begin{abstract}
In the shift towards human-centered manufacturing, our two-year longitudinal study investigates the real-world impact of deploying Cognitive Assistants (CAs) in factories. The CAs were designed to facilitate knowledge sharing among factory operators. Our investigation focused on smartphone-based voice assistants and LLM-powered chatbots, examining their usability and utility in a real-world factory setting. 
Based on the qualitative feedback we collected during the deployments of CAs at the factories, we conducted a thematic analysis to investigate the perceptions, challenges, and overall impact on workflow and knowledge sharing.

Our results indicate that while CAs have the potential to significantly improve efficiency through knowledge sharing and quicker resolution of production issues, they also introduce concerns around workplace surveillance, the types of knowledge that can be shared, and shortcomings compared to human-to-human knowledge sharing.
Additionally, our findings stress the importance of addressing privacy, knowledge contribution burdens, and tensions between factory operators and their managers. 

\end{abstract}



\maketitle

\section{Introduction}
An organization's success hinges on leveraging the knowledge found within the minds of its employees, customers, and suppliers - a process known as Knowledge Management (KM)~\cite{becerra2014knowledge,Nonaka.1995}. Knowledge management is increasingly recognized as a vital discipline that involves creating, sharing, and using an organization's knowledge. As such, understanding what leads to a successful Knowledge Management System (KMS) is an important area of research~\cite{chang2012factors, becerra2014knowledge}.

The importance of KM has motivated efforts to use technology, such as AI assistants, to support human knowledge work, a research field with over 60 years of history. Yet, the success of these systems was often held back by the knowledge acquisition bottleneck~\cite{wagnerBreakingKnowledgeAcquisition2006c}---capturing and maintaining human knowledge is highly resource intensive. On top of this, social factors such as the trade-off between perceived benefit and effort for the knowledge sharer pose additional challenges~\cite{hoffman2022retrofitt}. Domain knowledge can be complex, situational, and ephemeral, making it difficult to express and store, and it can quickly become outdated~\cite{tanLiveCaptureReuse2006}. Yet, cyber-physical systems, such as Cognitive Assistants (CA), show promise in supporting knowledge management~\cite{hoffman2022retrofitt}. \emph{In this work, CAs are AI systems that can interact conversationally to support cognitive tasks, for example, in supporting factory operators to share knowledge and learn about problems with the production machines}~\cite{KernanFreire2022ACognitive,leyerPowerWorkersEmpowering2018}.

Due to the complexities of KM, Natural Language Processing (NLP) techniques have had limited use in KM until recently, especially when handling unstructured texts such as manufacturing issue reports~\cite{mullerDigitalShopFloor2021,edwards2008clustering,orucc2020semantic}. Prior work on using assistant systems to share knowledge in factories has focused on using NLP technologies to deliver instructions efficiently but not the capturing of knowledge. For example, \citet{hoerner_using_2022} presented an assistant system for tacit knowledge sharing in manufacturing but still relies on human experts for the initial knowledge capture. There are examples of gamified experiences to elicit (manufacturing) knowledge; however, these have not been validated in practice~\cite{fenoglio_tacit_2022,balayn2022ready}. Therefore, there is a knowledge gap in understanding the real-world challenges and risks when using NLP technologies such as conversational AI for both knowledge capture and sharing on the factory floor.

Recent advances in NLP, such as Large Language Models (LLM) like the GPT-4~\cite{openai2023gpt4} or Claude family~\footnote{\url{https://www.anthropic.com/claude}---last accessed \today}, have increased the potential of CA systems. A foundational LLM can offer various functions, such as answering general knowledge questions, refining text, and checking logic. Additionally, the information contained in the foundational models can be extended by providing it with context material, a process called Retrieval Augmented Generation (RAG)~\cite{lewis2020retr}. LLMs are capable of more sophisticated, flexible, and human-like reasoning than previously possible with AI. These attributes could help overcome issues with agential AI during work support, such as not considering the social and context aspects of knowledge sharing~\cite{cabitza2021need,shneiderman2022human}.

Operating a modern production line is knowledge-intensive, dynamic, and socio-technical in nature~\cite{kotthaus2022nego}. As such, it is important to follow a human-centric approach when deploying new AI tools~\cite{shneiderman2022human}. Work by~\citet{hoffman2022retrofitt,morikeInvertedHierarchiesShop2022}, has shown how the social and hierarchical structures in manufacturing play a key role in knowledge sharing practices. Yet, one of the challenges of conducting research in manufacturing is the restrictions management imposes due to safety or a need for oversight, such as limited time to interview operators or requiring that a manager always be present. Despite the potential to influence how open the study participants are in discussing work matters, this is often unavoidable as factory management is vital to coordinating any research in factories~\cite{cheonWorkingBoundedCollaboration2022}. Thus, the (unfiltered) perspective of factory operators is poorly represented in literature.

In tackling these gaps, this work explores the human, organizational, and practical factors surrounding deploying CA systems in the real-world factory context from both the operator's and managers' perspectives, leading to the following research question:  


\textit{What are factory operators' and management perceptions of the impact and socio-technical risks and challenges of using cognitive assistants for knowledge sharing?}\label{RQ1}


    
    

The CSCW field has long contributed to Knowledge Management, taking a socio-technical and practice-based perspective~\cite{ackerman2013sharing}. We aim to expand on this body of knowledge by examining the latest generation of AI knowledge sharing tools for manufacturing, (LLM-powered) CAs. Over a period of two years, we deployed CAs in four phases, ranging from technology probes simulated by researchers to voice assistants connected to a live production line to LLM-powered chatbots. Using feedback from system evaluations and responses from semi-structured interviews conducted at two factories, we present findings from a hybrid deductive/inductive thematic analysis of \textit{N=40} operators' and management's perceptions of CAs.

Through our longitudinal study, we aim to contribute to a better understanding of the challenges and impact of using CAs in manufacturing. Thus supporting future research and implementation of CAs for knowledge sharing that are both effective,e and sensitive to the socio-technical context of operating complex systems in factories. Our contributions can be summarized as follows:

\begin{enumerate}
    \item We provide a deeper understanding of the opportunities, challenges, and risks associated with using (LLM-powered) CAs by factory operators from both operator and management perspectives.
    \item We provide insights on the tensions between factory operators and management regarding technology-facilitated knowledge sharing.
    \item We present design guidelines for effective (LLM-powered) CAs in manufacturing settings.
\end{enumerate}




\section{Background and Related Work}\label{background}
This section provides an overview of current research in Knowledge Management (KM) and the utilization of technology, specifically Cognitive Assistants (CA), to enhance knowledge sharing in organizational settings. It discusses the challenges and advancements in KM practices, the significant role of technological interventions in facilitating knowledge acquisition, dissemination, and application, and the socio-technical aspects that influence implementing these systems in manufacturing environments.

\subsection{Knowledge Management}

Organizational KM involves leveraging the knowledge of employees, customers, and suppliers~\cite{becerra2014knowledge}. This entails processes to acquire, manage, and share knowledge within an organization to achieve a competitive advantage~\cite{Nonaka.1995}. In practice, the core tenet of KM is to ``save content, it will be useful later''~\cite{abubakarKnowledgeManagementDecisionmaking2019}. For this work, CAs were deployed to support knowledge sharing between factory operators. In the following sections, we will outline the key processes surrounding knowledge sharing when facilitated by a digital system such as a CA, namely, knowledge acquisition, sharing, and application.

\textbf{Knowledge acquisition}---involves eliciting knowledge, explicating it, and formalizing it for later use~\cite{cookeVarietiesKnowledgeElicitation1994}. Knowledge can be elicited once it has been created in the mind of an employee. \textbf{Eliciting} involves extracting knowledge from domain experts or sources, often through interviews, surveys, observation, or analysis of existing documentation. Elicitation can uncover both explicit and tacit knowledge, understanding not just what is known but also how it's applied in practice. In this work, we focus on explicit knowledge that can be readily expressed in words as opposed to tacit knowledge that can not~\cite{nonaka2009knowledge}.

Once knowledge has been elicited, it needs to be made explicit. \textbf{Explication} involves articulating the knowledge clearly and understandably. This may involve defining terms and organizing information into coherent structures. Explication helps ensure that knowledge is comprehensible to others and can be effectively communicated.

Finally, during \textbf{formalization}, knowledge is structured and represented in a formal language or framework. Formalization makes knowledge more precise and facilitates analysis and manipulation by digital systems. This might involve creating models, rules, ontologies, or other formal knowledge representations. Formalization helps facilitate (AI) reasoning and decision-making based on the captured knowledge.

\textbf{Knowledge sharing}---is defined as the process of transferring knowledge between individuals, groups, or organizations~\cite{alaviReviewKnowledgeManagement2001}. Once knowledge has been acquired, it can be shared with others who can benefit from it. Traditionally, this can be done through various channels such as documentation, training sessions, workshops, seminars, meetings, or digital platforms. While many of these strategies aim to stimulate face-to-face knowledge sharing, organizations are also interested in systems that formalize the knowledge. This ensures that the knowledge persists beyond employees' memory and facilitates sharing at scale and asynchronously. In this work, we explore using CAs that act as an intermediary to help efficiently share newly created knowledge.

\textbf{Knowledge application}---is when knowledge is used to solve problems and make informed decisions to profit the organization~\cite{probstManagingKnowledgeBuilding2000}. This involves analyzing the situation, identifying relevant knowledge, and applying it effectively to address the issue at hand. In the context of this work, the CAs contribute primarily to the process of identifying and retrieving relevant knowledge. Analyzing the situation and taking action is still largely up to the human operator.

Knowledge management systems typically face many human, organizational, and technical challenges, such as employees hiding their knowledge to maintain (perceived) power or inefficient knowledge sharing practices~\cite{becerra2014knowledge}. 
Addressing these challenges, organizations have developed their own methods and systems to facilitate knowledge sharing. 


\subsection{Knowledge Sharing Systems}

Designing KS systems requires consideration of a broad range of factors, often intertwined with social and human elements. Organizations across various sectors have developed their own dedicated KS systems. For instance, the `Inside IBM' project used AI, information systems, and user-centered design to aggregate IBM's accumulated product support knowledge into a single system~\cite{masseyReengineeringCustomerRelationship2001}. Siemens employs TechnoWeb for social collaboration, which leverages a combination of social networking tools, collaborative platforms, and knowledge repositories to facilitate the sharing and management of expertise across the organization~\cite{lakhani2013open}. TechnoWeb integrates features such as discussion forums, expert directories, and document sharing to create a dynamic and interactive environment for knowledge exchange. NASA uses the Lessons Learned Information System (LLIS)\footnote{\url{https://llis.nasa.gov/}---last accessed \today} to document and share project experiences. LLIS employs a structured database approach, where lessons are systematically categorized, indexed, and made searchable to ensure that critical insights from past projects are easily accessible for future missions~\cite{liebowitz2002nasa}. This system includes detailed metadata tagging, cross-referencing of related lessons, and a review process to validate the accuracy and relevance of the information. Furthermore, the enhanced version of LLIS that the Goddard Space Flight Center uses employs an `active' push feature to recommend lessons that match the knowledge needs of specific individuals. \emph{These diverse systems illustrate the various methods employed to manage and leverage knowledge within organizations, ranging from searchable lessons learned and social collaboration tools to active knowledge recommendations and systematic documentation processes.}

Several factors contribute to the success of KS systems, as exemplified by the cases of Google, Toyota, and Xerox's Eureka project described below. In the case of Google, they emphasize psychological safety for KS~\cite{duhigg2016google}. Toyota---perhaps KM's most renowned manufacturing success story---employs network-level knowledge-sharing processes. These processes effectively engaged members in sharing critical knowledge, deterring free-riding behavior, and minimizing the obstacles to locating and obtaining valuable knowledge~\cite{dyer_creating_2000}. The Eureka project at Xerox showcases the successful use of technology to gather, validate, and share best practices across their customer service engineers~\cite{brown2000balancing}. By involving customer service engineers throughout the design process and encouraging local champions to support adoption and feedback, Xerox contributed to the project's success~\cite{bobrow2002community}. Ultimately, the Eureka project is said to have saved Xerox \$100 million~\cite{bobrow2002community,brown2000balancing}. \emph{Overall, these cases highlight the importance of minimizing the burden on knowledge contributors, ensuring high knowledge quality, and involving end-users throughout the design process to address their needs.}

\subsection{Integrating AI in Knowledge Sharing Systems}
While Eureka was an early example of an online knowledge base that service engineers could manually record and retrieve reports about machine problems, expert systems were the first widely used AI systems to reason over a knowledge base to support human decision-making. Using AI assistants for KM has progressed significantly from early expert systems in the 1970s, such as Dendral~\cite{LINDSAY1993209} and MYCIN~\cite{shortliffe_computer-based_1975}, which were designed to answer questions in specific domains based on rules defined by domain experts. In recent years, the improvements in natural language processing and deep learning have given rise to more sophisticated AI assistants, such as IBM's Watson~\cite{high2012era}.

KS systems incorporating AI and Internet of Things (IoT) technologies are transforming KS within manufacturing environments~\cite{casillo2020}. Yet, despite the promise of these technologies to enhance KS systems, several socio-technical hurdles persist. For instance, significant resources are required to develop and maintain knowledge bases~\cite{cullen1988knowledge}, and data quality issues often undermine efforts to automatically uncover knowledge from existing unstructured issue reports~\cite{edwards2008clustering,mullerDigitalShopFloor2021,orucc2020semantic}. Similar challenges have faced the use of NLP in healthcare, such as IBM Watson's Oncology Expert Advisor in 2016, which searched through `unstructured' physician notes to help suggest treatments. Yet, Watson's NLP failed to effectively process the jargon, shorthand, and subjective comments or pick up the nuances that human physicians could, demonstrating the challenges in handling human-generated texts~\cite{stricklandIBMWatsonHeal2019}. These challenges are socio-technical because they involve both technical aspects, such as data quality and the capabilities of NLP in processing it, and human aspects, such as the effort put toward documentation, which inhibits the data's utility for KS~\cite{8258120}. \emph{With recent advancements in NLP, such as the advent of LLMs such as GPT-4~\cite{openai2023gpt4}), it may be possible to overcome some of the hurdles of leveraging unstructured human documentation faced by older systems. Additionally, in tackling the human aspect, there is a knowledge gap in designing KS systems that facilitate the acquisition of human knowledge while balancing the (perceived) effort of knowledge authoring such that factory operators are consistently engaged in contributing useful knowledge.}

Foundational LLMs do not contain context-specific knowledge---for example, how to fix a machine in a specific factory---they possess extensive general knowledge and reasoning abilities~\cite{zhao2023survey}, enabling them to excel in processing complex information~\cite{jawahar-etal-2019-bert}, generate insights and reasoning~\cite{wei2022emergent}. While an LLM can be trained from the ground up for a specific domain, such as medicine, this is typically cost-prohibitive for individual organizations. Even so, using a domain-specific LLM would have two key challenges: (1) reliance on outdated information from their pre-training data, and (2) potential inaccuracies in factual content, a phenomenon termed ``hallucination''~\cite{bang2023multitask, zhao2023survey}. To mitigate these issues and maximize the utility of LLMs in specialized, knowledge-intensive tasks, techniques such as fine-tuning, chain-of-thought~\cite{weiChainofThoughtPromptingElicits2022} few-shot prompting~\cite{brown2020, gao-etal-2021-making}, and Retrieval Augmented Generation (RAG)~\cite{lewis2020retr} can be adopted. RAG involves retrieving relevant information from a knowledge base before generating an appropriate response for a human. RAG is an efficient way to harness the advanced NLP and reasoning abilities of LLMs to answer people's questions on domain-specific topics without needing to train a new model from scratch. Furthermore, the availability of source material---the information retrieved from the knowledge base---improves transparency and enables fact-checking. \emph{As such, foundational LLMs used in conjunction with knowledge bases hold significant potential to enhance organizational KM.}

\subsection{Cognitive Assistants for Factory Operators}
In recent years, research into AI assistants in manufacturing has increased, taking advantage of recent advancements in NLP and the proliferation of IoT to support decision-making, operationally efficiency, and training ~\cite{casillo2020,hannolaAssessingImpactSociotechnical2020b,rasBridgingSkillsGap2017,colabianchiHumantechnologyIntegrationIndustrial2023,hoerner_using_2022,kiangalaExperimentalHybridCustomized2024a,castaneASSISTANTProjectAI2023a,tao_self-aware_2019}). Although the social and human aspects are largely missing~\cite{hannolaAssessingImpactSociotechnical2020b}, they have been receiving increasing attention as demonstrated by the recent paradigm of Industry 5.0 where human well-being is key~\cite{brequeIndustrySustainableHumancentric2021,xuIndustryIndustryInception2021}. In the following sections, we discuss research on AI assistants in manufacturing, the use of knowledge bases, knowledge acquisition from humans for knowledge sharing, and the associated socio-technical challenges.

\subsection{AI Assistants}
In factories, human operators collaborating with AI assistants can be viewed as a socio-technical system consisting of the operator, the operator's tasks in the factory, and the technology enabling the assistant~\cite{maedche2019ai}. AI assistants can help operators fix problems with the machines or set them up optimally by enabling efficient information retrieval, remote machine control, decision support, and sharing knowledge~\cite{dhuiebDigitalFactoryAssistant2013,bergerHowDesignValuebased2023,castaneASSISTANTProjectAI2023a,liBringingNaturalLanguageenabled2022}. To enhance their capabilities, AI assistants can be integrated into factory systems, such as control and monitoring systems for production machines, scheduling systems, and sensors. Some systems make use of the available data to perform predictive maintenance using Machine Learning (ML)~\cite{wellsandtHybridaugmentedIntelligencePredictive2022a} or NLP to extract and represent knowledge from texts~\cite{naqviHumanKnowledgeCentered2022a}. To interact with operators, most AI assistants use conversational AI---a chat or voice interface---to provide efficient and natural ways for operators to communicate.

Recently, manufacturing companies have cautiously adopted advanced AI assistants that use LLMs. For instance,~\citet{mercedes2023} integrated ChatGPT into vehicle production to improve error identification, quality management, and process optimization. This allowed quality engineers to simplify complex evaluations and presentations of production data through dialogue-based queries. Furthermore,~\citet{xia2023autonomous} showcased how in-context learning and the injection of task-specific knowledge into LLMs can improve the planning and control of production processes. \emph{However, systems like these rely on an up-to-date and detailed knowledge base, a major hurdle for any KMS.}

\subsection{Knowledge-intensive AI Assistants}
The knowledge bases and rules that inform AI assistants in manufacturing are usually static, having been defined by a domain expert during development~\cite{trappeyVRenabledEngineeringConsultation2022,rooeinChattingProcessesDigital2020,casillo2020,wielandRuleBasedManufacturingIntegration2016}. Some systems use live data but are focused on presenting the status of production systems and perform (limited) reasoning over the data~\cite{liBotXAIbasedVirtual2021a,reisVirtualAssistanceContext2022,mantravadiUserFriendlyMESInterfaces2020j,pavlovCaseStudyUsing2020,baldaufHumanInterventionsSmart2021}. For example, \citet{casillo2020} presents a chatbot for training new operators based on a predefined curriculum, obtaining promising results. \citet{baldaufHumanInterventionsSmart2021} deployed smartwatches that notified operators of machine errors, citing the importance of concise information display. Additionally, the chatbot demonstrated by \citet{trappeyVRenabledEngineeringConsultation2022} utilizes VR to deliver knowledge in response to FAQs (frequently asked questions). \emph{Overall, existing research on AI assistants focused on understanding how to deliver knowledge or present (live) information to operators but not acquire knowledge from operators.}

\subsection{Cognitive Assistants: Acquiring and Sharing Human Knowledge}
In the context of this study, \textbf{Cognitive Assistants (CAs) are an advanced type of AI assistant that interacts conversationally to acquire knowledge from humans and share it with other humans, supporting decision-making.} Cognition refers to the mental processes of acquiring and comprehending knowledge~\cite{anguloCognitive2023b}, which the CA aims to support. To do so effectively, a CA can use several technologies, including dialogue management, NLP, context awareness, databases, and ontologies. While deploying CAs at a production line appears novel, we discuss notable works that evaluated (parts of) the underlying technologies and techniques in the following sections.

\textbf{Acquiring knowledge using conversational AI and/or mobile devices} has been explored and shown to be promising. For example,~\citet{fenoglio_tacit_2022} introduced a role-playing game involving virtual agents, human experts, and knowledge engineers to refine knowledge graphs that were algorithmically generated. The authors raised several important ethical concerns regarding the processing of personally identifying data, deciding to avoid audio and video recordings, and other concerns regarding misuse of employee monitoring. \citet{hoerner_using_2022} built a digital assistance system to support operator troubleshooting processes on the shop floor using captured knowledge. While they allowed operators to suggest edits, \emph{knowledge engineers still performed the initial knowledge capture process.} In contrast, \citet{hannolaAssessingImpactSociotechnical2020b} deployed smartphone applications for factory operators to take notes and videos of solutions to production line problems themselves, thus enabling other operators to access newly created knowledge. \emph{While this work highlighted the potential of facilitating KS through digital technologies, it did not investigate the application of LLMs and/or conversational AI.}

Examples outside of manufacturing demonstrated the successful use of conversational AI to crowdsource knowledge acquisition, namely, a game to elicit knowledge from crowdworkers~\cite{balayn2022ready} and a context-aware chatbot that simultaneously fulfills information needs while acquiring new knowledge~\cite{curious.2017}. A key consideration for these systems is ensuring the quality of acquired knowledge, thus employing rating or validation mechanisms. \emph{Although these conversational AI systems were not deployed in a manufacturing environment, they demonstrate the potential of using conversational interactions to efficiently acquire knowledge.}

\subsection{Socio-technical Challenges}
While many KMS have been deployed in manufacturing, the human potential for knowledge sharing and the associated social factors have been largely overlooked~\cite{hannolaAssessingImpactSociotechnical2020b}. Work in factories is socio-technical rather than purely technical, deterministic activities~\cite{kotthaus2022nego,muchaIndustrialInternetThings2018}. While it is clear that digital tools can facilitate knowledge sharing among users, for example, by enabling factory operators to post comments under existing work instructions to facilitate discussion about best practices~\cite{ludwig20173D}, these tools can have many unforeseen social effects and barriers. For example, introducing relatively simple tools, such as a machine repair ticketing system, can shift the balance of control towards or away from operators depending on who is given privileges to create, view, and modify tickets~\cite{kotthaus2022nego}. Furthermore, operators omitted their valuable informal knowledge when asked to create learning material collaboratively in favor of the standard working procedures~\cite{weinert2022designing}. Additionally, requiring operators to use smartphones to document their work can be perceived as controlling~\cite{hannolaAssessingImpactSociotechnical2020b}. Thus, these factors emphasize the importance of considering social factors when introducing new digital knowledge sharing practices, which are generally overlooked when designing AI assistant systems~\cite{cabitza2021need}.

To tackle the complex social challenges discussed above, it is crucial to look beyond the technical aspects when developing AI tools for knowledge sharing. Although not social in nature, considering human-computer interaction factors such as avoiding information overload and ensuring the information needs are met on an individual level are also key~\cite{hoffman2022retrofitt}. Indeed, recent work on assistance systems for sharing tacit knowledge in factories that had focused on technological aspects emphasized the importance of comprehensively evaluating the social, psychological, and organizational implications~\cite{hoerner_using_2022}. The crucial importance of socio-technical factors is what motivated us to holistically reflect on the feedback we received from operators and managers while deploying CAs, culminating in this study.

\section{Real-World Case Studies in Factories}\label{method}

\subsection{Approach and Context}
We conducted a hybrid deductive/inductive thematic analysis of 251 comments collected from operators and factory management over two years of CA evaluations at two factories in two European countries. The factories manufactured detergents on production lines that could rapidly switch between over a hundred detergent types depending on customer demand. The factories operated 24/7 with three eight-hour shifts per day, and each production line was typically manned by two operators. The operators reconfigure, operate, and resolve issues with the production lines, all knowledge-intensive tasks, especially considering the frequent production and quality issues. We evaluated four CAs of varying levels of sophistication and functionality, from technology probes simulated by researchers to CA systems connected to a live production line, allowing us to explore many facets of the operators's experiences. This research was approved by the factory operators' councils and our institution's human research ethics committee.

\subsection{Systems, Protocols, and Participants}

The development of the CAs used in this study was initiated according to objectives set by the factory management: to increase production performance and reduce training time by supporting (tacit) knowledge sharing amongst operators and training operators to follow standardized procedures. The factory management had observed significant disparities between operator shift performance, poor adherence to standard working procedures, and inefficient human-human knowledge sharing practices. Despite the top-down initiation of the project, we involved the operators throughout the development and evaluation of the system to ensure that we were also meeting their needs and values. The evaluated functions are listed in Table~\ref{table:capabilities} and participants in Table~\ref{table:participants}. In the following sections, we describe four phases of data collection, including the capabilities of the CAs, data collection protocols, and participant information.

\begin{table}[ht]
\centering
\begin{tabular}{c|l|c|c|c|c}
\textbf{No.} & \textbf{Capability} & \textbf{Phase 1} & \textbf{Phase 2} & \textbf{Phase 3} & \textbf{Phase 4} \\
\hline
1  & Speech input                                & \checkmark &            & \checkmark & \checkmark \\
2  & Speech output                               & \checkmark &            & \checkmark &            \\ 
3  & Text input                                  &            & \checkmark & \checkmark & \checkmark \\ 
4  & Text output                                 &            & \checkmark & \checkmark & \checkmark \\ 
5  & Capture issue handling knowledge            & \checkmark & \checkmark & \checkmark & \checkmark \\ 
6  & Share issue handling knowledge              &            & \checkmark & \checkmark & \checkmark \\ 
7  & Automatic validation of captured knowledge  &            &            &            & \checkmark \\ 
8  & Human approval of captured knowledge        &            &            &            & \checkmark \\ 
9  & Capture product knowledge                   &            & \checkmark & \checkmark &            \\ 
10 & Rate captured product knowledge             &            & \checkmark & \checkmark &            \\ 
11 & Approve shared product knowledge            &            & \checkmark & \checkmark &            \\ 
12 & Share product knowledge                     &            & \checkmark & \checkmark &            \\ 
13 & Graph machine data for systematic reflection &   &            & \checkmark &            \\ 
14 & Machine vision for context-aware dialogue       &   & \checkmark & \checkmark &            \\ 
15 & Upload new documents to the knowledge base           &   &            &            & \checkmark \\ 
16 & Answer predefined FAQs                               &   & \checkmark & \checkmark &            \\ 
17 & Provide relevant standard work instructions          &   & \checkmark & \checkmark &            \\ 
18 & Answer queries using factory documentation           &   &            &            & \checkmark \\
19 & Real-time explainable machine settings advice & & & \checkmark & \\
\end{tabular}
\caption{CA capabilities evaluated per phase.}
\label{table:capabilities}
\end{table}

\begin{table}[ht]
    \centering
    \caption{Participant Data}
    \begin{tabular}{c|c|c|c|c|c|c}
    No. & Role & Factory & Phase & Test duration (min) & Prior Phases & Gender \\
    \hline
    1 & Operator & 1 & 1 & 60-180 & No & Male \\
    2 & Manager & 1 & 1 & 0 & No & Male \\
    3 & Operator & 1 & 1 & 60-180 & No & Male \\
    4 & Manager & 1 & 1 & 0 & No & Male \\
    5 & Operator & 1 & 1 & 60-180 & No & Male \\
    6 & Operator & 1 & 1 & 60-180 & No & Male \\
    7 & Technician & 1 & 1 & 0 & No & Male \\
    8* & Mixed & 1 & 2 & 10 & No & Mixed \\
    21 & Manager & 1 & 3 & 15-30 & Phase 2 & Male \\
    22 & Operator & 1 & 3 & 15-30 & No & Male \\
    23 & Technician & 1 & 3 & 15-30 & No & Male \\
    24 & Operator & 1 & 3 & 15-30 & No & Male \\
    25 & Operator & 1 & 3 & 15-30 & No & Male \\
    26 & Manager & 1 & 3 & 15-30 & Phase 2 & Male \\
    27 & Manager & 2 & 4 & 15-30 & No & Female \\
    28 & Manager & 2 & 4 & 15-30 & No & Female \\
    29 & Manager & 2 & 4 & 15-30 & No & Male \\
    30 & Manager & 2 & 4 & 15-30 & No & Female \\
    31 & Operator & 1 & 4 & 15-30 & No & Male \\
    32 & Operator & 1 & 4 & 15-30 & No & Male \\
    33 & Manager & 1 & 4 & 15-30 & Phase 2,3 & Male \\
    34 & Operator & 1 & 4 & 15-30 & No & Male \\
    35 & Manager & 1 & 4 & 15-30 & Phase 1,2,3 & Male \\
    36 & Operator & 2 & 4 & 15-30 & No & Male \\
    37 & Operator & 2 & 4 & 15-30 & No & Male \\
    38 & Operator & 2 & 4 & 15-30 & No & Male \\
    39 & Operator & 2 & 4 & 15-30 & Phase 3 & Male \\
    40 & Operator & 2 & 4 & 15-30 & Phase 1 & Male \\
    \end{tabular}
    \label{table:participants}
\end{table}

\subsubsection{Phase One: Technology Probe during Production Line Operations}
\begin{figure}[ht]
    \centering
    \includegraphics[width=0.6\columnwidth]{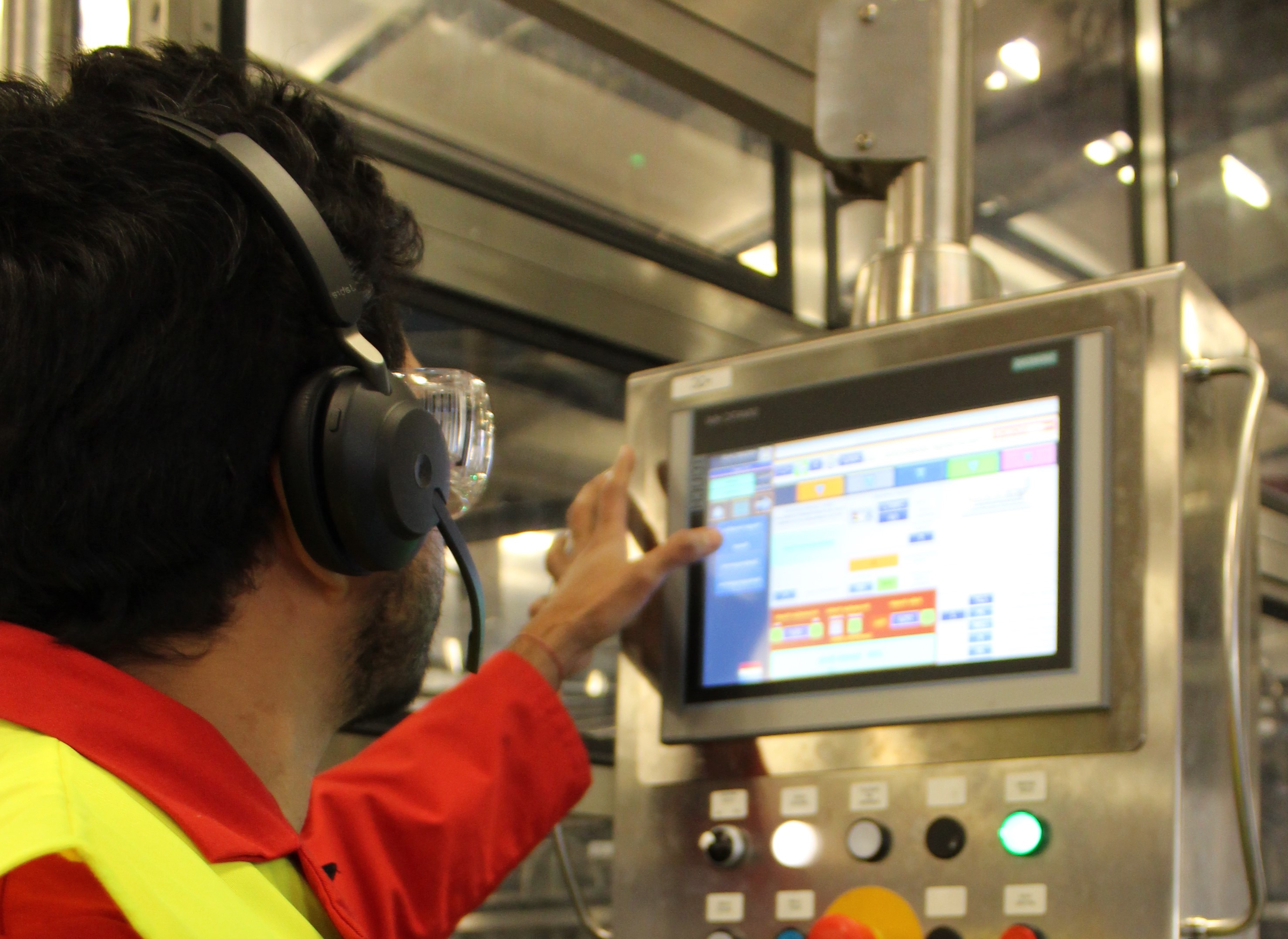}
    \caption{Headset used at the production line for data collection during phase 1}
    \Description{TBD}
    \label{fig:headset}
\end{figure}
In preparation for the first evaluation phase (Jan. 2021-May 2021), we asked the operators (\textit{n=2}) and their managers (\textit{n=2}) to describe the current knowledge management practices at the factory. Together, we defined the information needed to represent an operator's knowledge regarding issue handling, namely, the symptoms, machine component(s) associated with the symptom, type of issue, location, current task, solution, and root cause analysis. During the first phase of evaluations, we used a technology probe~\cite{boehner2007hci} to evaluate the user experience of using a CA for capturing knowledge at the production line. Over three days, in May 2021, we asked factory operators to share how they solved issues with a CA being simulated by a researcher. Participants used a Bluetooth headset to interact verbally whilst at the production line. The researcher simulated the CA using a protocol representing state-of-the-art NLU and dialogue management. The protocol followed a form-filling conversation during which the operator was asked to describe the issue according to the abovementioned representation. If the operator missed specific information or used pronouns (e.g., \emph{it}) to describe components, the researcher would ask for clarification. Data was collected over three days, and six (\textit{N=6}) male operators participated. Additionally, the user study was partially observed by two managers and a technician. Throughout the user study, researchers recorded observations by the managers and operator feedback regarding the operators' experiences and challenges during interactions with the CA, resulting in 18 (\textit{$n_c$=18}) comments relevant to using CAs in factories. Key findings from the user study include the negative impact of factory noise on voice interaction, discomfort from wearing headsets for prolonged periods, and the need for shorter interactions when eliciting information from operators.

\subsubsection{Phase Two: CA with a Simulated Production Line}
Having learned the importance of efficient and reliable interactions from the first phase of evaluations, we developed a CA that used live data from a simulated production line and human location tracking using stereoscopic cameras\footnote{\url{https://www.stereolabs.com/products/zed-2}---last accessed \today.} for context awareness. We built a section of a production line in the lab, including human-machine interfaces and a computational model that simulated the machine's behavior to develop the location tracking capabilities in a stable environment. Then, we installed the stereoscopic cameras at the factory. Context awareness enabled the text-based CA to facilitate efficient user interactions by prefilling user responses as buttons depending on their proximity to machines using the human tracking system and live data from the machines. For example, during knowledge sharing about issue handling or machine setup, the CA would use the live data to suggest error codes or machine components and ask the operator to confirm. These shortcuts enabled faster and more reliable interactions.

As voice interaction was unreliable due to factory noise and frequent use of factory-specific jargon, we opted to develop a text-only CA for phase two. The CA was built using version 1.0.6 of RasaX\footnote{\url{https://legacy-docs-rasa-x.rasa.com/docs/rasa-x/1.0.x/}---last accessed \today.}, an open-source framework for building context-aware conversational agents. The assistant used an action server to connect to additional services, most importantly a Neo4j\footnote{\url{https://neo4j.com/}---last accessed \today.} knowledge graph for storing the captured knowledge. We conducted three focus group sessions, totaling 11 (\textit{n=11}) factory operators and managers, two (\textit{n=2}) of which identified as female, and nine (\textit{n=9}) as male. The sessions consisted of the following four activities: watching a video presenting the simulated production line, a video presenting the CA and how to use it, a structured user test that involved completing eight tasks with the CA, and finally, a semi-structured user test where participants were free to experiment with the CA, evaluating its capabilities in supporting operators. The tasks for the user test focused on retrieving and providing information to the CA aided by live data from the simulation but did not require the participants to act upon the retrieved information; for example, we instructed the participants to ``Find instructions on how to perform a prerun'', ``Record the current filling machine settings'' and ``Ask for help with this problem: `Symptom: The filler is foaming. Error code: 33. Product: 7508226.') At the end of each activity (e.g., after completing all eight tasks), participants were asked to write post-it notes on their experience and how it could be improved. Additionally, we facilitated group discussions for which a second researcher took notes. The anonymously collected (post-it) notes resulted in 55 (\textit{$n_c$=55}) comments that were relevant to this study. We excluded comments that were not directly related to the CA, for example, if they were solely related to the realism of the production line simulation. The participants from phase two are referenced collectively as P8* as we did not track comments to individuals.

\begin{figure}[ht]
    \begin{subfigure}{0.49\columnwidth}
        \includegraphics[width=0.9\columnwidth]{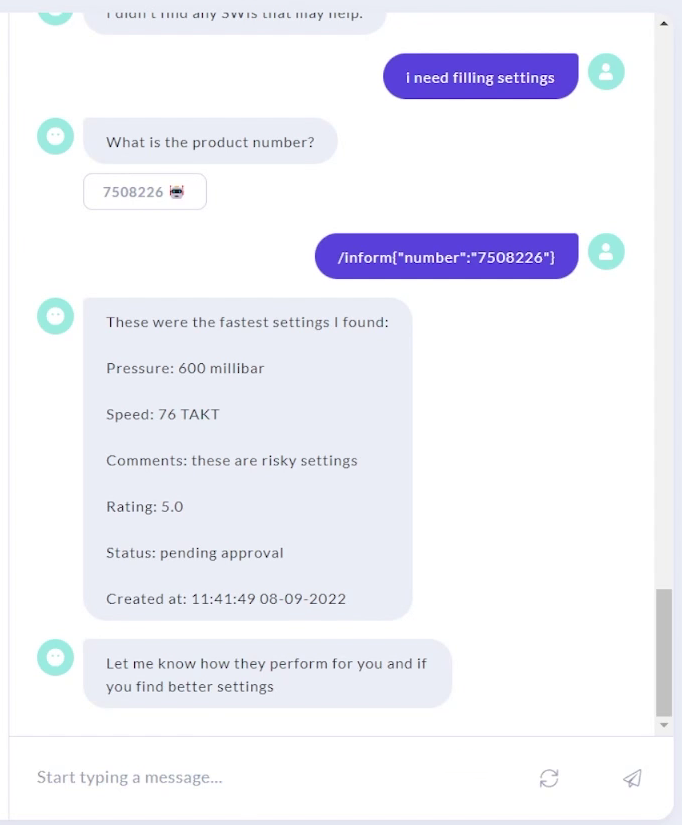}
        \caption{}
        \Description{UI}
        \label{fig:phase2UI}
    \end{subfigure}
    \begin{subfigure}{0.49\columnwidth}
        \includegraphics[width=0.9\columnwidth]{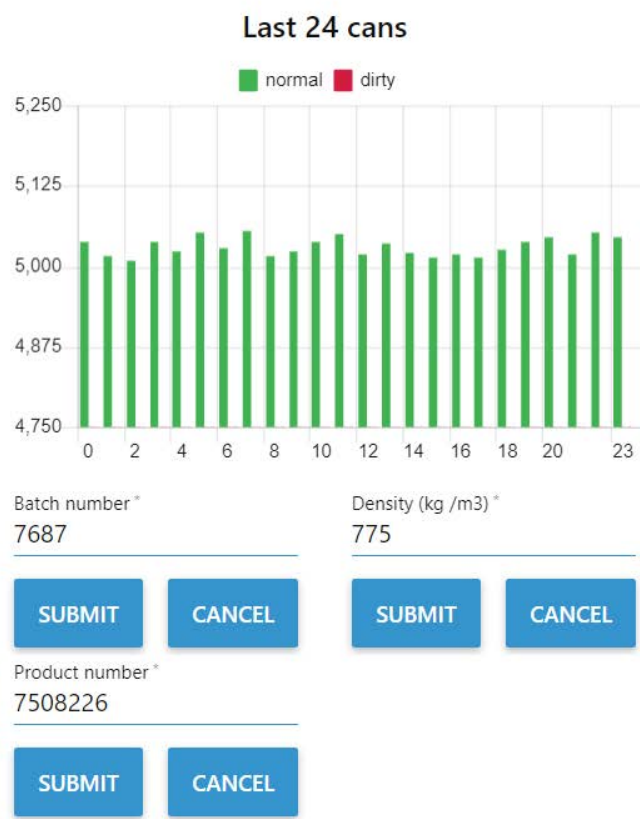}
        \caption{}
        \Description{The image shows a graphical user interface for the weight checker. A bar chart displays the weight of the last 24 cans, fields for error codes, and input fields to enter information such as product numbers.}
        \label{fig:simulation_weightchecker_view}
    \end{subfigure}
    \caption{(a) The Rasa X user interface from phase 2 and (b) A simulated ``weight checker'' user interface that was connected to the assistant during phase 2. Each bar represents the weight of a canister that was filled with detergent. The ``weight checker'' weighs each canister and stops the production line if the weight falls below a specified threshold.}
    \label{fig:simulation}
\end{figure}

\subsubsection{Phase Three: CA with Partial LLM Integration Connected to the Production Line}
During this phase, we evaluated a CA capable of voice and text interaction with access to live machine data from the production line. In addition to context-aware user interactions, the data was used to provide ML recommendations and generate statistics and graphical representations of the production line performance.

As factory operators faced many problems during an 8-hour shift, it was challenging to capture all relevant new knowledge in a timely manner. Furthermore, when the factory operators were asked to look back on their shifts and prepare a summary for the next shift, it was challenging to remember all relevant events and information. To support reflections on the shift, we integrated a function to display statistics and visualizations of production performance over time. This served as an overview and memory prompt for the operators to look over.

Factory management had observed large discrepancies between the production speeds achieved by operators. Indeed, many operators were not aware of the speeds achieved by their colleagues for each product and what explained the difference. We trained a Random Forest (RF) classifier on historical data to provide recommendations in an effort to support operators in achieving higher production speeds. The output consisted of speed adjustment recommendations, namely increase, decrease, or no action. The recommendations were accompanied by a textual explanation based on the model features used, for example, ``I recommend increasing speed as the last canister weight was 123 grams over the threshold and the delta to the historical maximum speed of the product is 15 TAKT''.

The most significant change for phase three was the integration of an LLM-powered Retrieval Augmented Generation (RAG) function to answer user queries using the GPT-3.5 API (version gpt-3.5-turbo-0301\footnote{\url{https://platform.openai.com/docs/models/gpt-3-5}---last accessed \today.}) and a corpus of shared operator knowledge. The CA, which used the Rasa framework\footnote{\url{https://rasa.com/docs/rasa/}---last accessed \today.}, was accessible via an Android application on a smartphone based on the Mycroft companion app\footnote{\url{https://github.com/MycroftAI/Mycroft-Android}---last accessed \today.}. At this stage, we demonstrated the capabilities of the CA to five (\textit{n=5}) operators and three (\textit{n=3}) managers. All participants identified as male. The operators were given the opportunity to use the CA at the production line as they saw fit (e.g., to record and retrieve information about recent issues), whereas the managers were asked to complete five tasks in an office setting. We asked the managers to complete the following tasks as if they were operators: save and retrieve machine parameters and associated tips, save and retrieve a solution to a problem, generate production statistics for the past 12 hours, and finally, pose a question about a problem using the LLM-powered response generator. Following this, we conducted semi-structured interviews regarding their perceptions of its knowledge sharing capabilities, user interactions, benefits, risks, and suggestions for improvements. We extracted relevant statements from the interviews, resulting in 60 (\textit{$n_c$=60}) comments for this study.

\begin{figure}[ht]
    \begin{subfigure}{0.49\columnwidth}
        \includegraphics[width=0.9\columnwidth]{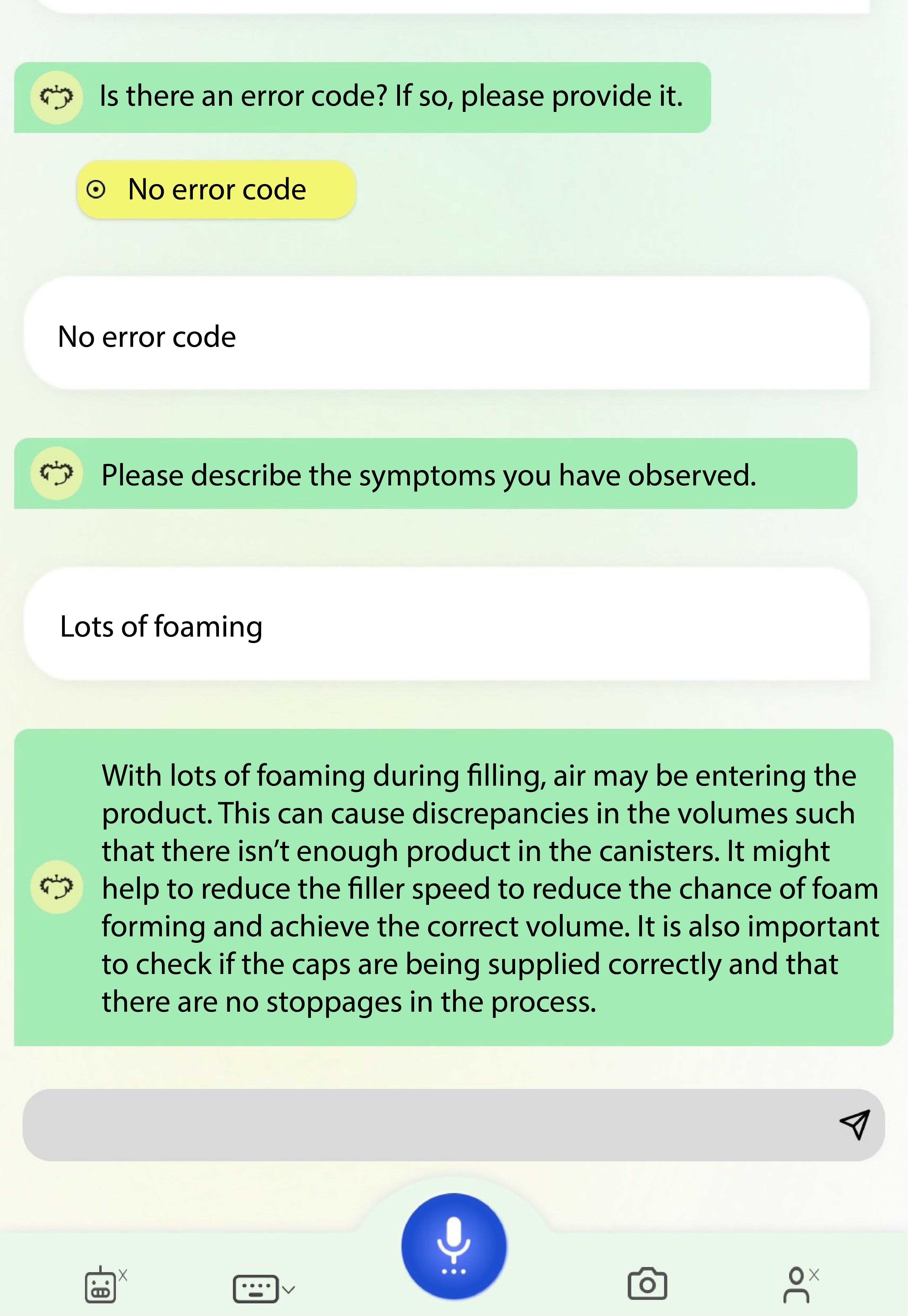}
        \caption{}
        \Description{Android user interface for phase 3 and 4}
        \label{fig:androidLLM}
    \end{subfigure}
    \begin{subfigure}{0.49\columnwidth}
        \includegraphics[width=0.9\columnwidth]{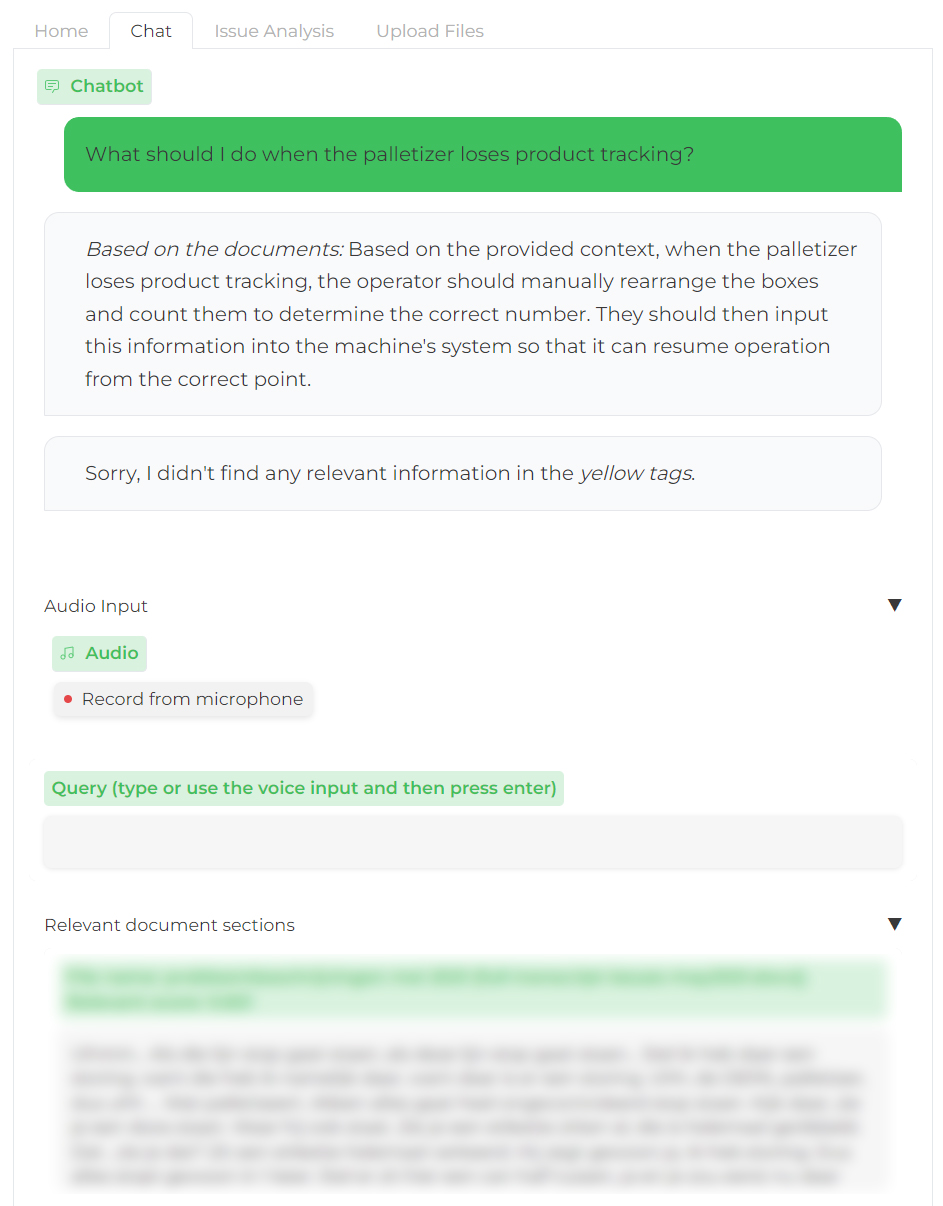}
        \caption{}
        \Description{gradio}
        \label{fig:gradio_interface}
    \end{subfigure}
    \caption{(a) Android user interface for phase 3 and 4 and (b) The query tab of the LLM-powered CA from phase 4}
    \label{fig:gradioLLM}
\end{figure}

\subsubsection{Phase Four: LLM-powered Cognitive Assistant}
During the final evaluations, we tested the LLM-powered functions exclusively, using the gpt-3.5-turbo-0613 model API\footnote{\url{https://platform.openai.com/docs/models/gpt-3-5}---last accessed \today.}, LlamaIndex\footnote{\url{https://docs.llamaindex.ai/en/stable/}---last accessed \today.} for the response generation and Gradio\footnote{\url{https://www.gradio.app/}---last accessed \today.} for the browser-based frontend. The LLM-powered system introduced the following new functions: RAG for user queries, a digital form for submitting issue analysis reports, automatic validation of submitted root-cause analysis reports, human validation of submitted root-cause analysis reports, and a function to upload new operating documentation.

The RAG feature enables factory operators to quickly get advice for issues or procedural issues they are facing by chatting with the CA. The CA retrieves information from two separate knowledge bases: standard operating documentation and operator-submitted issue analysis reports. We opted to keep these separate in the CA's response, so it is clear to the operator what the source of information is (See Figure~\ref{fig:gradio_interface}). Additionally, the operators can click on a drop-down menu to inspect the reference material directly.

In another tab of the interface, operators can fill out and submit issue analysis reports. We followed the standard practices of the factories by including fields for location, issue description, (root-)cause, and solution. As operators often struggled to find the root cause of issues, we included optional fields to conduct a 5-why analysis for root-cause identification~\cite{serrat2017five}. Upon completion of the report, an LLM prompt is used to automatically validate the logic of the reasoning steps and consistencies with existing knowledge in the knowledge base. Finally, an employee with adminstrator powers can approve the report, inserting it into the knowledge base for future use.

We evaluated the LLM-powered CA with three (\textit{n=3}) operators and one (\textit{n=1}) manager from the first factory, and four (\textit{n=4}) managers and five (\textit{n=5}) operators from the second factory. The evaluation in the first factory was conducted in person at a production line, with one researcher coordinating the evaluation and conducting the interview and a second researcher recording audio and taking notes. Conversely, the evaluation in the second factory was conducted asynchronously without the presence of a researcher. The operators and managers from the second factory were provided with an online survey that specified the tasks and open-ended questions that were used to extract the comments. The tasks for both factories aimed to evaluate the knowledge capture and sharing capabilities of the CAs with recent issues from their work. For example, we instructed the participants to ask for help with recent problems they faced on the production line. Similarly, we asked them to record a solution and their thought process for a problem they recently faced. After completing the tasks, the open-ended questions for both factories included the following topics: the perceived benefits, risks, adoption barriers, operator-management relations, and suggestions for improvements, resulting in 118 (\textit{$n_c$=118}) comments.

\subsection{Hybrid Deductive/Inductive Thematic Analysis}
After several rounds of qualitative data collection over two years, we conducted a hybrid deductive/inductive thematic analysis with two independent coders as described by ~\citet{fereday2006demonstrating} and recently applied by ~\citet{martinez-maldonado_lessons_2023}. The analysis was conducted based on 251 (\textit{$n_c$=118}) comments we extracted from interviews, survey answers, and post-it notes that were collected during phases one through four. All comments related to the user experience, social, technical and organizational factors of the CA deployments at the factories were included. Conversely, unintelligible comments were excluded. Complex statements comprising multiple standalone arguments or observations were divided into separate comments to facilitate sorting and analysis. If the other part(s) of the original statement provided relevant context for the divided comment, this was included in regular brackets.

For the thematic analysis, the first level involved a deductive approach to define themes closely related to the research questions and literature review, resulting in the following six themes: (1) Impact on Work Experience, (2) Optimizing Knowledge Sharing, (3) Adoption and Change Management, (4) Privacy, Safety, and Ethics, (5) Usability and User Experience and (6) Technical and Operational issues. The second level followed an inductive approach to identify subthemes by two independent coders. Finally, the subthemes were collaboratively revised to reach a consensus, resulting in 20 subthemes. At this stage, we resolved all intercoder discrepancies. At both levels, we allowed comments to be sorted into multiple themes and subthemes to accommodate a holistic understanding of participants' perspectives on real-world issues.


\section{Operator and Management Perceptions}\label{analysis}
\subsection{Impact on Work Experience}
\subsubsection{Improved Information Retrieval}
The comments suggest that the system positively impacts information retrieval within the organization. The CA was perceived to provide immediate answers to doubts about machine operation, making access to company documents more user-friendly (P30), and improving the flow of information (P28). Furthermore, participants saw it as a tool that modernizes the factory (P27). However, one concern is that operators may benefit more from traditional information retrieval systems (P40).

\subsubsection{Useful for Problem-Solving}
Many participants (\textit{n=9}) are positive about the usefulness of the system and specific functions (P22, P24, P25, P36, P30, P31, P32, P33, and P34), and some participants, such as P25, make a distinction between some functions being useful and others not. Specifically, the system provides greater speed in carrying out some small tasks (P29), it has a well-set-up logical control function (P30), and helps solving problems quickly (P8*, P24, P25, and P37) as stated here ``I find it nice that you can look back. What did we do last time with the same problem? That, I think, is a significant advantage.'' by P24, and ``[the CA] can be a great tool in problem solving in the day-by-day activities.'' by P8*. However, there is a disadvantage of wasting time looking for a solution to a problem if it is not reported in the system's history (P29).

\subsubsection{Effective Knowledge Sharing}
Several benefits regarding knowledge sharing are mentioned by the participants, namely that it is useful to transfer knowledge from experienced to novice operators (P31), ``The system also allows for the communication of tacit knowledge between teams.'' (P33), it can also help individuals recall knowledge of how they previously solved issues (P24, P32) and experts' knowledge can now persist and is always available (P34). These comments show that participants generally see it as a mechanism for experts to share their knowledge with novices and as a memory augmentation for themselves.

That said, there are concerns that receiving information from the system could be less quick, effective, and comprehensive than asking the expert staff present (P27). This is likely true, assuming that there is an expert present with the relevant knowledge. However, our observations at the production line and statements by operators, such as ``A few hours are lost every week on previously solved issues.'' (P33) and P34 stated that operators ``Lose about half an hour per day over one team per line.'' suggests that information on previously solved issues was not readily available.

Whereas management believes in standardizing approaches so every operator operates at the highest achievable performance, some operators do not think this is feasible. One operator in particular, P25, was skeptical of the benefits of knowledge sharing as stated here, ``We always adjust to the specific product characteristics and line conditions.'', ``We often have different issues.'' and ``Because everyone has their own approach, some run slower, and others have their own techniques.''. In other words, their work is so complex that the `best practice' is highly dynamic and dependent on context factors and individual strategies. This reflects experienced operators' pride in the strategies they developed, which may hinder the effectiveness of knowledge sharing. Conversely, P22 mentioned that ``It would be useful to know if and how other operators have produced at higher speeds.'' (P22), suggesting they would be interested in learning from each other. This reflects a split among operators regarding the perceived usefulness of knowledge sharing.

\subsubsection{Training Time}
Opinions are divided on whether training time will be reduced, with P32 stating that ``[the CA] will address staff shortages and training time.'' whereas P31 stated that training will not directly be reduced but independence will be increased. Others, such as P33, were unsure whether training time would be impacted but expect improvements as the technology matures. Generally, participants thought that novice operators would benefit most (P8*, P22, P24, P25, P31, P34, and P38); for example, P22 stated, ``This is very nice, yes, especially for beginner [operators].'' and P25 stated, ``It's useful for newcomers who might not have as much product and line knowledge.'' These statements support the idea that independence can be improved and training time can be reduced.

\subsection{Optimizing Knowledge Sharing}
\subsubsection{Value of Efficient Knowledge Sharing}
The comments highlight the importance of operator knowledge at work, confirming the importance of capturing this knowledge. This know-how is not captured in manuals or documentation; for example, ``An operator poured water over one of the canisters in the filling machine when the line had stopped to get it moving again.'' (P6) and ``We adapt to the product's characteristics, and some products require slower operation due to foam production.'' (P25). Yet, existing mechanisms for capturing this knowledge were largely unused, as stated by P2: ``Because the problems are so poorly documented in [current issue reporting system], we don’t have a good overview of the problems.'' P25 also describes how they develop their own strategies: ``Because everyone has their own approach, some run slower, and others have their own techniques.'' and the necessity to adapt to changing context: ``Yes, sometimes we have products that change their speed, or the machine is modified, allowing faster operation.'' This partially explains why existing tools for documenting knowledge are quickly outdated and resource-intensive to maintain, as discussed in Section \ref{analysis:kmaintenance} below. Overall, these points demonstrate the high potential value of an efficient knowledge sharing mechanism.

\subsubsection{Knowledge Representation}\label{analysis:representation}
There were some criticisms of the generic nature of (some of) the information provided (P39). Related to this, there were suggestions for improvement, such as more detailed, precise, and updated information (P27, P37, P40) and more specific instructions (P40). This suggests that knowledge representation should be comprehensive and precise, focusing on providing detailed instructions or information where necessary. Indeed, participants were concerned that some knowledge would be too complex for the system; for example, ``Problems at the production line can be more complex and have complex solutions.''; ``Mechanical problems can be hard to describe. pictures might be better.''; and ``It will be challenging to match similar problems using text only as people can describe them in many different ways.'' (P8*). Technically, this was a significant challenge when using keyword-based search and intent-based assistants; however, semantic search and LLM-powered RAG can handle divergent phrasing much better. Other operators believed a simple approach would be sufficient; for example: ``Error code and description would be enough information to save.'' (P24), and P22 reiterates that ``Error codes are the single most important factor to match to existing issue solutions.''. However, P25 points out that ``Sometimes there are random issues that don't have specific fault codes.'', demonstrating a need for alternatives. These wishes reflect a desire to minimize interaction time with the system. Therefore, a tension exists between the perceived effort required to document knowledge and the value of that knowledge for sharing purposes.

\subsubsection{Knowledge Maintenance}\label{analysis:kmaintenance}
The comments discuss the necessity and challenges of knowledge maintenance from several perspectives. P40 states that ``If the system is not kept updated, it becomes useless.'' and P24, whilst reflecting on previous tools for KM, mentioned that ``The biggest challenge is keeping it updated. It takes so many man-hours, and you just don't have those.''. Additionally, participants raised concerns about not having sufficient data, such as P8*, who stated that ``[The system] might not give good recommendations if it's only half full of knowledge.'' and ``[The system] needs more data.''. There are also concerns about how to share knowledge clearly if there are many (potentially conflicting) entries in the knowledge base to share (P8*).

Some resistance to participating in the upkeep of knowledge is evident as operators reported that uploading documents is not their responsibility (P40). There were also concerns about how the database will be connected and updated with instructions and documents (P8*), and the need for a proper process for approving knowledge updates to maintain high quality, as stated here by P8*: ``We will need to have a proper process for the settings change confirmed by an engineer.''. Comments on the existing KM situation at the factory portray the challenges in keeping the knowledge updated, as discussed in section \ref{adoption_practices}.

\subsection{Adoption and Change Management}
\subsubsection{Operator Support and Assistance}\label{analysis:adoption:support}
The comments suggest that operator support and assistance are crucial for adopting the CA. P29 highlights the importance of awareness and training for operators, which could potentially hinder adoption if not adequately provided. P37 also emphasizes the need for operators to know how to use the system correctly to improve adoption. 

\subsubsection{Operator Acceptance and Adoption Challenges}
The comments reveal several challenges to operator acceptance and adoption of the CA. Managers are concerned that ``If two or three bad experiences occur, then [operators] might be discouraged from using it.''. Some participants refer to the need for full adoption; for example, P30 suggests that adoption could be hindered if all employees don't accept the system, indicating a need for broad buy-in. Furthermore, P38 implies that older operators' usage could be a potential barrier to adoption, and P34 also points out that ``[User acceptance] Depends on digital literacy of operators.'', suggesting that age and technological proficiency may play a role in adoption. However, P34 clarifies that they have no concerns about adoption. P35 reveals privacy concerns as a potential obstacle to adoption, with one operator refusing to participate due to discomfort with human tracking, as discussed in Section \ref{analysis:privacy}. This remains a critical challenge for adoption, especially if rogue operators sabotage the system. Other potential barriers to adoption include poor performance (\ref{analysis:performance}) due to lack of details and out-of-date knowledge (\ref{analysis:representation}), poor accessibility (\ref{analysis:design}), lack of cultural importance for knowledge sharing (\ref{adoption_practices}), and a high perceived effort to record knowledge (\ref{analysis:representation}).

A few participants referred to failed attempts to introduce new documentation tools; for example, P23 points stated: ``We bought a bunch of tablets 7 years ago with the intention of using them for documentation, but they were never adopted.'' and that current documentation tools are poorly used (see Section \ref{adoption_practices} for more details.) Although we do not know the reasons behind the failed adoption of tablets for documentation, it does reflect how challenging introducing a new KM system can be, and why both management and operators need to commit.

\subsubsection{Management and Operator Perspectives}
Using a CA to share knowledge among operators is largely motivated by the factory management's observations that this was not naturally occurring effectively despite their attempts to stimulate it. This is reflected in the following statement by P3, an operator: ``Management has accused me of not sharing my knowledge with others. However, it was due to a miscommunication about what the problem was about.''. The comments generally indicate a range of perspectives between management and operators regarding knowledge sharing practices. Some operators---especially those who were very distrustful of management---strongly believe that knowledge is already sufficiently shared and do not appreciate the suggestion that they are hiding it. Regarding documentation, some operators stated that operating the production line is too intensive, leaving insufficient time for reporting. From the management's perspective, they observe operators who are frequently idle and not proactive in sharing their expertise. Regardless of the truth, we observed that the careful introduction of a knowledge sharing tool such as a CA is key to reconciling the differing perspectives. For example, by framing the CA as a support tool for operators to avoid the impression that experienced operators are being forced into sharing their knowledge. 

Indeed, many of the managers believed that the CA would need to be carefully introduced to avoid rejection from the operators. P29 suggests the need to sensitize operators and explain the system's advantages, implying a potential gap in understanding or perception between management and operators. P38 and P39, both operators, suggest that perspectives may vary depending on the individual, with some being indifferent to the differences between management and operators. P1 reveals a desire for operators to be heard by upper management, whereas others are satisfied with the influence operators already have and feel insulated from the desires of (upper) management through, for example, the workers' council (P32). That being said, several operators mentioned that the management implements new tools without fully committing. This has led to distrust toward management regarding whether they would commit to the CA. Conversely, the management believes that the tools they introduced work sufficiently well, and the operator's laziness is the problem. Overall, our experience is that most operators were open to new knowledge sharing tools introduced by management as long as they were involved in the process, the benefit was clear, and the management invested sufficiently in delivering a reliable and useful tool.

Regarding the knowledge sharing interactions with the CA, we observed differences in requirements between the operators and management. Management was mostly concerned with collecting sufficiently detailed knowledge and ensuring operators used the tool to encourage standardized best practices. Conversely, the operators were concerned with ensuring the interactions were as short as possible, minimizing the required details and only using the tool when absolutely necessary. While (upper) management tended to focus on an idealistic view of capturing every detail of an operator's knowledge about a problem, the operators tended to seek out the minimally required detail for describing and retrieving problem solutions. Concurrently, the vision of (upper) management was that all operators follow a single `best practice', whereas some operators preferred to follow their own intuition and personal strategies, at least when it came to tuning the parameters of the production line. When it came to solving critical problems that stopped production, operators were more open to learning from others. Overall, we recognized a stark difference in perspectives between management and operators, both with valid concerns. These factors demonstrate the complex socio-technical challenges of introducing a tool for sharing knowledge.

\subsubsection{Integration with Existing Systems and Practices}\label{adoption_practices}
The comments suggest that integrating the CA with existing systems and practices is a key consideration. P1 revealed that some operators use WhatsApp for internal communication, suggesting that operators were not satisfied with official communication channels. That being said, the operators still used the company-issued phones to request help from technical services or expert operators, many of whom were more senior. The official policy for resolving problems at the production line consisted of several steps of escalation, beginning with the present operators trying to solve the problem themselves. If unable to solve the problem, operators can request help from other (expert) line operators, shift leads or technical services. If a CA were introduced, it could be positioned as the first escalation step, allowing (novice) operators to elicit support before needing to call on their fellow operators, who are likely busy with other tasks. Most operators supported the idea of sharing their knowledge to support each other in this way, especially less experienced operators and those involved with training operators. However, it remains to be seen if the operators would be willing to maintain the effort for a prolonged period of time.

At the production line, we noticed a pattern where operators frequently skipped or improperly executed tasks they deemed non-essential. Operators revealed that some new procedures are not always followed, such as autonomous maintenance procedures (small maintenance tasks that operators are expected to routinely conduct): ``[The autonomous maintenance] cards are frequently turned around without the task being completed.'' (P3). Furthermore, there is evidence that operators do not properly input information into existing documentation systems, as stated by P2: ``Because the problems are so poorly documented in [issue reporting system], we don’t have a good overview of the problems.'' (P2) and ``Existing digital report system is frequently unused.'' (P5). When we asked the operators why they were not using the issue reporting system, we frequently heard that it was unreliable, suggesting that the operators did not see the point of putting in effort to use something that could not be relied upon. Furthermore, the upkeep of standard working procedures has not been performed as mentioned by P24 here: ``No, no, but in fact, the current standard working instructions are from 2003 or 4 when we started with it. It has become such a tangled mess. We really want a different system for it. It's all still in Excel.'' All of this points to a lack of investment in knowledge management systems from the management and a lack of engagement in responsibilities that are perceived as non-essential to operators, possibly reflecting a work culture where operators are disinclined to conduct tasks beyond the operation of the production line. Interestingly, P33 mentioned that ``Previous non-digital version existed and was popular but more cumbersome.'', indicating that engagement with a previous knowledge sharing system used to be higher but was hindered by its unwieldiness. While the introduction of a CA alone will likely not change the underlying culture directly, it could reduce the perceived effort in documentation and improve the perceived benefit of documentation.

\subsection{Privacy, Safety and Ethics}
\subsubsection{Concerns over Operator Privacy and Data Handling}\label{analysis:privacy}
The comments reveal mixed opinions regarding user privacy and data handling. Some users, as indicated in comments P31, P32, P33, and P34, do not express any privacy concerns, with P33 stating, ``No [privacy concerns as] most of the data isn't personally identifiable.'' However, P35 reveals a contrasting perspective, where an operator refused to participate due to privacy concerns and was uncomfortable with human tracking and covert management oversight. P34 suggests a balanced approach, advising the provision of individual profiles for operators but also cautioning about privacy. This suggests that while some users are indifferent or unconcerned about privacy, others are acutely aware and concerned about how their data is handled and monitored.

\subsubsection{Security Measures Against Misinformation or Malicious Use}\label{analysis:malicious}
There are some concerns regarding the security risks if the system's advice is incorrect, as stated by P27: ``Risks: if the responses are not adequate, you risk safety.''. Safety becomes especially critical in inherently dangerous industries such as detergent factories, the context of this study. This raises questions regarding who is accountable for mistakes, especially if the system relies on knowledge shared by other humans, and what level of accuracy is acceptable for the system if human safety is at stake. Furthermore, P33 mentioned a ``Small possibility exists that users may be malicious and enter falsehoods but unlikely.''. The above points support the necessity of mechanisms to safeguard the quality of the knowledge, such as knowledge input approval procedures, checking for conflicts with existing knowledge, and identifying outdated knowledge, many of which could also be supported by AI (see also Section \ref{analysis_tech_data}). Furthermore, framing the system as a supporting tool, not a decision-maker, could help remind operators that they are still accountable, as suggested by P34: ``Will be useful as a supportive tool but not a solution''.

\subsubsection{Operator Autonomy and Empowerment}
Participants emphasize the importance of operator autonomy and see the CA as a support tool, as stated by P33: ``[It will benefit operators] as long as it remains a support tool and operators remain the main actor in the work.''. P34 suggests giving each operator their own profile to aid knowledge management, indicating a desire for operator autonomy and empowerment. However, the same comment also cautions about privacy. Indeed, as discussed above in Section \ref{analysis:privacy}, P35 mentioned that an operator refused to participate due to privacy concerns about the camera system, underscoring the importance of operator autonomy. On this topic, operator opinions are divided, as P5 states, ``No issues with having cameras and would rather that [management] invest in enough cameras to cover the entire line instead of half-measures.'', indicating a desire for management to fully commit to a solution that can support operators. Throughout the process, we respected operator autonomy by involving the workers' council in the design process. They approved the work we conducted, and everyone was given the opportunity to ask questions and discuss concerns. However, despite formally approving the deployment of a camera system, at least one operator sabotaged the installment, indicating disagreement indirectly, as mentioned by P21: ``[Operators] have placed a sticker over the camera, blocking its view.'' (see Figure \ref{fig:sticker} below). This highlights an ethical dilemma that organizations and researchers may face when introducing AI systems that track humans: operators may not feel safe to oppose the introduction of new tools openly.

\begin{figure}[ht]
    \centering
    \includegraphics[width=0.6\columnwidth]{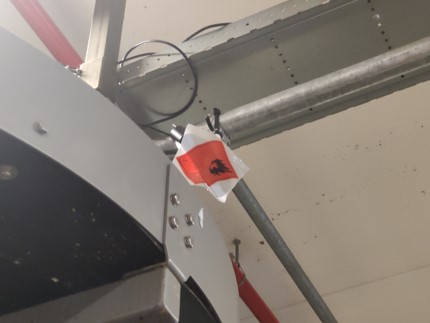}
    \caption{A sticker used to block the stereoscopic camera for anonymous human tracking that overlooks the production line.}
    \Description{TBD}
    \label{fig:sticker}
\end{figure}

\subsection{Usability and User Experience}
\subsubsection{User Interface Design and Accessibility}\label{analysis:design}
Operators highlighted the importance of a user-friendly, accessible system that caters to diverse user needs. Operators expressed a desire for a more intuitive information display, with one suggesting to ``Display the attachments after the main text for easier reference.'' (P28). Similarly, several suggestions highlight the shortcomings in the systems NLU, for example, requests for recognizing keywords to provide relevant suggestions and streamline communication (P8*). The use of visual aids was also highlighted, with suggestions to ``use more pictograms and pictures to speed up understanding.'' (P8*). Language localization was another aspect that users found important, with requests for system versions in other (local) languages. Mobile accessibility was appreciated, but concerns were raised about the use of personal devices, with one user questioning, ``Are we not going to force people to use their phone for that?''(P24) These comments highlight the need for a system that is easy to navigate, visually intuitive, available in local languages, and providing company devices.

\subsubsection{System Performance and Intelligence}\label{analysis:performance}
The topic of reliable and efficient user interaction is frequently mentioned. Whilst nine (\textit{n=9}) participants are positive regarding the system's natural language understanding, eight (\textit{n=8}) participants expressed concerns or areas for improvement, highlighting the challenges for user satisfaction. Suggestions include the ability to understand user queries based on keywords (P8*), including autocomplete in the chatbot to speed up the performance, and better NLU when asking for assistance during machine troubleshooting (P8*). There were also calls to load more (detailed) data and knowledge into the system (e.g., P27) to improve performance. Despite these concerns, \textit{n=9} participants provided positive feedback on the system's performance and usability, suggesting that it is generally well-received and effective in its current form. Two participants appreciated the response speed; for example, P29 stated: ``Quick in responses.''. Others highlighted the desire for ``More consistent performance.'' (P33). The emphasis on reliable, speedy performance is logical, considering one of the main perceived advantages of the system is to speed up problem solving as stated by P37: ``Advantage: speed in problem solving.''



\subsubsection{User Training}
The participant suggests training the operators to use the system effectively; for example, P8* stated that ``Probably a short training or demonstration would be needed to be able to use it properly.'' This demonstrates that despite the system supporting natural language interaction and being able to explain how to use it, participants still value formal training to improve user experience. Furthermore, P36 stated that ``Advantages in the future with generational change with adequate training, no particular difficulties.'' and P8* also mentioned that once operators get used to the system, it becomes easier, suggesting that with some help, the system was easy to use.

\subsection{Technical and Operational Issues}
\subsubsection{Network and Connectivity Issues}
The comments indicate significant network and connectivity issues that affected the operations. For instance, P35, and P23 highlight that Wi-Fi and 4G internet within the factory were unreliable. This disrupted communication and data transfer, crucial for smooth user interactions with the assistant as it relied on an internet connection. Additionally, P21 points out firewall issues that hindered access to certain resources. This demonstrates the importance of a reliable and well-managed Wi-Fi network within organizations considering deploying connected systems such as CAs.

\subsubsection{Technical Limitations and Troubleshooting} 
We faced several challenges integrating the system with factory systems. P35 suggests that coordinating with the (3rd party) factory IT was challenging due to their frequent unavailability or preoccupation with other tasks. Furthermore, P21 mentions a technical issue where a PC was inaccessible, highlighting the limitation of coordinating with multiple technical parties to integrate systems.

\subsubsection{Dependence on Accurate and Complete Data Input}\label{analysis_tech_data} 

The comments highlight challenges related to the `cold start problem': providing benefits and driving engagement if the system's knowledge base is empty. Indeed, during evaluations, P8* indicated that the system needs more data, confirming this issue. Motivating operators to share their knowledge is a considerable challenge, especially initially. Furthermore, once the system has reached a point where it provides perceptible benefits, it will still need to be maintained and kept up-to-date. Factory management also required that the CA include a validation step so that an expert operator can approve all incoming knowledge. Furthermore, P24 emphasizes that the biggest challenge they faced with previous KM systems was keeping them updated due to the significant man-hours required. This suggests a need for more efficient data input methods or automation to ensure the system has accurate and complete data. Additionally, a mechanism is required to remove and update existing knowledge as the machines and products can change over time; P25 states, ``Yes, sometimes we have products that change their speed or the machine is modified, allowing faster operation.''. This underscores the importance of having an efficient mechanism for keeping a knowledge base up-to-date and approving incoming knowledge.

\section{Discussion}\label{discussion}
The domain of Computer-Supported Cooperative Work (CSCW) has evolved significantly over the decades, aligning with technological advancements, a focus on labor relations, and shifts in work practices. Originally, knowledge sharing in CSCW focused on centralizing expert knowledge through the `repository model' in the late 1980s. This model primarily relied on forming extensive databases that document expert insights. However, this strategy struggled with situational variability---knowledge is often highly contextual and not easily generalized across different scenarios. Furthermore, the requirement for experts to codify their knowledge into repositories imposed a significant authoring burden, detracting from their primary tasks and responsibilities.

Subsequently, the field shifted away from static repositories towards facilitating direct exchanges between individuals, aiming to capture the nuanced and dynamic nature of knowledge. However, scalability remained a challenge, and the effectiveness of these interactions hinged on the availability of both parties, which is not always feasible in a high-paced shift-based production environment.

Most recently, our research aligns with the paradigmatic shift towards cyber-physical systems as intermediaries for knowledge sharing, reflected in CSCW research~\cite{hoffmannCyberPhysicalSystemsKnowledge2019a}. These systems, particularly through advancements in AI and live data processing, aim to manage the complexity of real-time knowledge sharing. This leads to critical discussions in CSCW and manufacturing on how to share knowledge from a human to machine and visa versa~\cite{abubakarKnowledgeManagementDecisionmaking2019,schniederjansSupplyChainDigitisation2020,listarossettiIdentifyingIndustryTechnologies2023}. Although, still underrepresented, our work contributes to the small but growing body of CSCW work in manufacturing such as~\cite{gerbrachtWeAreNo2024,cheonWorkingBoundedCollaboration2022,hoerner_using_2022,hoffman2022retrofitt,morikeInvertedHierarchiesShop2022,wurhoferReflectionsOperatorsMaintenance2018}. Our work goes beyond prior CSCW work on knowledge sharing in manufacturing such as~\cite{hoffman2022retrofitt,hoerner_using_2022,hannolaAssessingImpactSociotechnical2020b} by learning from evaluating fully functioning CAs in real-world production environments and integrating recent technology advancements, namely LLMs.

Our findings suggest that while (LLM-powered) CAs effectively bridge gaps left by earlier paradigms, such as scalability and minimizing the authoring burden, they also introduce new dimensions to consider, such as the potential perceptions of surveillance, diminished work fulfillment, and the risks of overreliance on AI. Through this discussion, we situate our work within the broader CSCW knowledge sharing research, highlighting the challenges and lessons learned in the form of design guidelines (\textbf{G\#}).

\subsection{Cognitive Assistants Excel at Sharing the Solutions to Emerging Issues}

In our study, the CAs were perceived to be effective at sharing knowledge, especially when acquiring knowledge from experienced operators to support novices. Both management and operators recognized that emerging issues at the production line warranted rapid dissemination of solutions that currently did not happen effectively. As production lines become increasingly complex, digitized, and adaptable, operators spend a considerable amount of time and cognitive resources on error handling, in line with findings of prior work~\cite{wurhoferReflectionsOperatorsMaintenance2018}. Concretely, a few hours of downtime could be saved on the production line weekly if solutions to emerging issues are shared effectively. To maximize this benefit, it is crucial to ensure that information can be found quickly and the knowledge base can be easily updated or modified, leading to the following guidelines: \textbf{Design CAs to facilitate rapid information retrieval to support timely resolution of issues}~\guideline{} and \textbf{Design CAs to enable quick modification and additions to the knowledge base}~\guideline{}. Overall, participants universally agreed on the benefits of sharing solutions to issues such as machine malfunctions or problems caused by human error to resolve the issues quicker.

In contrast to sharing solutions for recurring problems, the topic of sharing knowledge about tuning the machines to achieve higher performance is controversial for some operators. This is especially true for those who are comfortable with operating at lower, stable speeds. These operators were skeptical about the benefits of knowledge sharing on machine fine-tuning as they apply unique strategies that others cannot easily adopt and are highly situated. Machine tuning is a complex and iterative process that is affected by many factors, such as ambient temperature and fluctuating raw material properties, making it more complex to share, aligning with observations by prior CSCW research~\cite{hoffman2022retrofitt}. On top of this, according to management, some operators attempted to hide their knowledge to maintain their position of power. As such, we observed resistance to sharing valuable knowledge on machine tuning for various reasons.

While some experienced operators see no benefit in learning from others concerning machine tuning, this contradicts the management perspective, who observed significant differences in the production speeds attained by different experienced operators and were trying to encourage them to learn from each other. 
Indeed, other operators still recognized the benefit of sharing some information about machine tuning, such as attainable target production speeds, the accompanying parameters, and some expert tips.
Ultimately, in this study, we shifted focus to supporting the operators during issue resolution instead of machine tuning, as the benefits were clear to all, and we wanted to keep the operators interested in the CA. Thus, we would advise to \textbf{collaborate with operators to identify the scenarios where sharing knowledge via a CA can result in significant and perceptible support for the operators}~\guideline{}.

\subsection{Considering the Authoring Burden of Knowledge and the Relevance for other Operators}
Regarding the effort of sharing knowledge, several participants brought up the challenge of motivating engagement from operators. While improving accessibility and keeping dialogues short can minimize the perceived effort of using the CA, the perceived benefit of sharing knowledge to help others remains a central barrier to adoption aligning with prior work~\cite{hannolaAssessingImpactSociotechnical2020b,hoerner_using_2022}. Looking at the factors that affect technology acceptance~\cite{davis1989perceived}, perceived benefit and perceived effort as opposing forces that will impact the attitude toward using the technology. External factors can also influence adoption, such as incentives, cultural norms, and voluntariness to help colleagues. In this study, we observed the strongest motivation to share knowledge among the `expert' operators who were already involved in training novice operators as part of their job. 

Going deeper into the tensions surrounding the balance of providing and receiving knowledge, we assume that more effort invested in eliciting and capturing knowledge will result in a more comprehensive knowledge base that can be used to help others. Understandably, the balance may shift depending on many factors, such as its potential impact, how time-critical it is, if it's something only one expert has discovered, how broadly applicable it is, how situational it is, how readily it can be shared, etc. For example, a solution to a recurring and high-impact problem that only one operator has discovered should be immediately captured and shared, even if it takes considerable effort. Additionally, the act of sharing knowledge is in tension with conducting the primary task of operating the production line, echoing previous CSCW work by~\citet{hoffman2022retrofitt}. It is also important to consider the ephemeral nature of knowledge as the complex system changes and new knowledge is discovered. As such, organizations could consider a prioritization framework to guide operators on how much effort should be invested in sharing their knowledge on a case-by-case basis. This prioritization could be adjusted computationally depending on frequently asked questions that the CA receives as an indicator of relevance or production urgency. \textbf{Thus, considering the burden of sharing knowledge on operators, it is important to consider the value of knowledge to prioritize what to share and when}~\guideline{}

Maintaining the temporal relevance and quality of the knowledge base is a key concern from both management and operator perspectives. The managers involved stipulated that all new knowledge entries should be approved by designated `expert' operators, similar to mechanisms deployed by~\citet{hoerner_using_2022}, providing some level of quality control. In this study, management was enthusiastic about using LLMs to validate the knowledge inputs by checking logic and consistency with existing knowledge. The validation output can be used by the operator to improve the report before submitting it for human approval.

Although not implemented in our work due to the anonymity policies we agreed with operators, other work has made the author of entries visible to others~\cite{hannolaAssessingImpactSociotechnical2020b}, which could support the validation and knowledge application process but also introduce potential bias' concerning the quality of specific operator's knowledge~\cite{yangWhenKnowledgeNetwork2019b}. Additionally, \textbf{the validity of entries should be continuously evaluated, for example, by enabling operator feedback or by identifying conflicts between entries and safeguarding against outdated knowledge}~\guideline{}. Overall, several strategies can be employed to maintain a high-quality knowledge base while considering the potential for bias, the burden on knowledge maintenance, and resulting improvements.

\subsection{Educating the Local Stakeholders is Key}
Operators who fundamentally disagree with the benefit of sharing some knowledge may avoid using the CA. Indeed, assuming that the CA can support operators in doing their job better, the operators that are open to using the CA and can effectively interact with it will be at an advantage. For new operators, this means they can become more independent of their human mentors more quickly, and more experienced operators can benefit from the collective knowledge of their colleagues. In contrast, the perceived value of the operators not using the CA may be reduced, for example, because they reject it fundamentally or have difficulties interacting with it. This highlights the importance of \textbf{considering accessibility and, where possible, encouraging full operator adoption, for example, by involving them throughout the development process and solving problems that are important to them}~\guideline{}.

During our study, many operators demonstrated a deep pride in the knowledge and expertise they apply in their work. It becomes critical, therefore, to frame technology, such as a CA, as a tool that supports rather than replaces their skill and judgment. This notion is in harmony with \citet{shneiderman2022human} who advocates for `humans in the group; computers in the loop', stressing that CAs should enhance rather than substitute human decision-making. \textbf{Emphasizing that operators are still responsible for their decisions even with CA support could help them stay critical and informed about the advice they receive from the system}~\guideline{}. Additionally, \textbf{communicating the strengths and weaknesses of the CA, such as being aware of potential `hallucinations' in responses from Large Language Models (LLMs) and the influence of leading questions, is crucial for productive interactions}~\guideline{}.

Yet, similar to prior research, we observed that workers' councils and operators often do not know enough about the technology to engage deeply in discussions concerning new AI systems~\cite{gerbrachtWeAreNo2024}. On top of this, upon reflecting on our own experience with the workers' councils, we recognize that they were only involved upfront during the project initiation when many details were still fuzzy. In contrast, after the project was approved, they were no longer involved. Considering the innovative and explorative nature of the project, it could have benefited from more continued input from the workers' council to safeguard the operator's interests and ensure they remained well-informed, as demonstrated by \citet{crabtree2017repacking,zhang2021an}. We also note that despite our efforts to explain the CA architecture and capabilities in our interactions with operators and posters we put up at the production, we observed that many operators only fully grasped the concept and its benefits once the full system was deployed. This could be attributed to the complexity and novelty of the system we were deploying, which perhaps warrants \textbf{additional effort when explaining the concept, for example, through tangible paper prototypes, making the process more accessible to the local stakeholders}~\guideline{}.

\subsection{Co-Evolving Local Infrastructure and Mitigation Plans for Technical Issues}
In addition to the social challenges, deploying CAs at the production line faces many practical and technical challenges that affect the system and our research. The factories in this study used numerous separate digital systems such as for planning, maintenance activities, quality control, productivity, and machine documentation. Many of these systems were considerably outdated and incompatible with each other. Resultingly, considerable effort was needed to connect them to the CA, sometimes requiring the involvement of third-party suppliers. For example, some of the underlying infrastructures, such as databases with API access, needed to be deployed before the research could proceed. As such, considering the effort involved, \textbf{relevant stakeholders should discuss whether connecting specific systems to the CA is worth it, considering factors such as added value, reliability, the investment needed, privacy implications, and upcoming system changes}~\guideline{}. As~\citet{edwards2010the} and~\citet{martinez-maldonado_lessons_2023} have highlighted, developing HCI infrastructures that adapt alongside work practices is not just beneficial but essential for the sustainable implementation of technological advancements in any setting.

The poor availability and quality of the data were compounded by poor (wifi and mobile) network stability in the factory, meaning the availability of live data could not be guaranteed. As a result, operators sometimes needed to wait 30 seconds for a response or could not receive the support they requested. Therefore, we advise that \textbf{companies should prepare reliable and robust data infrastructures before deploying CAs}~\guideline{}, for example, by providing reliable wifi and investing in consolidating (live) data in centralized databases on modern servers. Acknowledging that high-quality, live data will not always be available, designers and developers of CAs should \textbf{design cognitive assistants to be robust to missing or low-quality data}~\guideline{}. For example, mitigation strategies could include proactively asking the operator to provide critical information describing a problem (e.g., pressure levels) if the CA needs it to give reliable advice.



\textit{Accessible AI for Support, not Surveillance}
While production data and issue reports were frequently of low quality or out-of-date, operators were not very motivated to improve them. Several operators mentioned that the current system used for tracking productivity and issues tracking did not accurately measure production speed and sometimes went offline. The operators used this to justify why they did not use it to report issues or pay much heed to the reported production statistics. Then, when a manager tries to investigate why production was low for a shift, the operators claim that the data is inaccurate or does not represent the actual situation. This situation is similar to operator behavior observed by~\citet{morikeInvertedHierarchiesShop2022}, where operators deliberately entered incomplete data into a digital terminal so that management did not have a complete picture of their work. 
\textbf{Consider that operators may resist systems that provide managers with more accurate and timely information on production processes}~\guideline{}. Doing so gives the operators more control which can make their work more meaningful~\cite{braheStoryWorkingWorkflow2007,rossoMeaningWorkTheoretical2010}
Thus, understanding the origins of operator resistance to system changes is vital for designing CSCW and HCI solutions that will be adopted and used effectively.

Although most operators expressed concerns about privacy, particularly regarding surveillance by managers and being unfairly evaluated on work performance, the majority agreed to allow sensitive data collection, such as their location. The main requirement was that access was restricted, particularly from management, highlighting the necessity that the benefits of data collection should be clearly beneficial to operators and not just for surveillance, echoing findings of~\citet{dasswainSensibleSensitiveAI2024} that found information workers perceived tracking as pure surveillance when they were not presented any consumable insights. Surprisingly, the operators we interviewed went as far as to express favoring a more intrusive tracking system that could enhance the system's effectiveness over half-measures that did not benefit them. Thus, \textbf{any implementation of tracking technologies must carefully balance utility and privacy while including operators in the decision process}~\guideline{}.

A major challenge when involving operators in privacy discussions is catering to widely different perspectives. In our study, the CA has context awareness through an operator tracking system. The tracking system consisted of multiple stereoscopic cameras and a computer vision model to create an anonymous 18-point skeleton of operators along the production line. The tracking data was not personally identifiable, and factory employees could not access it. The workers' council and all the operators we spoke to approved the tracking system. Furthermore, all operators were informed about the tracking system via announcements and posters along the production line. Despite these efforts, the tracking system was sabotaged by an operator. Despite pressure from factory management to fix the tracking system and continue with the research, we interpreted the operator's actions as an implicit rejection of the tracking system and disabled it. Thus, beyond respecting the legal side of privacy, organizations will need to consider when to force the implementation of human tracking for CAs, especially if the majority of the workforce would prefer the additional utility that context awareness could provide, such as automatically capturing knowledge~\cite{yangWhenKnowledgeNetwork2019b} or suggesting responses.

Beyond context-aware suggested responses, several other factors can impact the accessibility of interactions with the CA at the production line. These factors include familiarity with CAs, difficulty reading text on a smartphone screen, physical access to the smartphone with the CA, noise that may impact the accuracy of speech-to-text and audibility of text-to-speech, (local) accents or jargon affecting speech-to-text and NLP, and existing knowledge on the abilities and limitations of AI technology, such as LLMs. Indeed, aligning the design of the CA with the needs of operators is a crucial area of research~\cite{fox2020worker}. The challenges of handling noise, accents, and local jargon when using speech-to-text remains an open point but beyond the scope of our work. In this study, we observed that providing (context-aware) suggested responses for the operators to click on instead of type responses was effective at improving accessibility. Furthermore, using advanced NLP technologies, such as LLMs, made the CA easier to use as it became more capable in understanding the operator's queries. The final version of the CA we deployed was also accessible in computer browsers instead of only through a smartphone app, improving accessibility for some of the older operators who had trouble using the smartphone. These integrations underline the \textbf{importance of designing technology that is adapted to real-world factory contexts and the capabilities of the operators}~\guideline{}.

\subsection{Implications for Researching and Deploying Cognitive Assistants in Factories}
Navigating the tensions between operators and management regarding AI-mediated knowledge sharing is key to the success of CAs in factories. As the management initiates the decision to bring in new technology like CAs, the operators can perceive this as a means of control and surveillance. As discussed by prior studies, the early and continued involvement of operators and workers' councils can facilitate the adoption of new technology but did not go into the specifics on how~\cite{cheonWorkingBoundedCollaboration2022,menewegerWorkingTogetherIndustrial2015}. The technology investigated in this study, (LLM-powered) CAs that support knowledge sharing, pose some unique challenges that warrant extra attention, such as the difficulty imagining the impact on work, the authoring burden on (expert) operators, focusing on knowledge that operators understand the benefits of dissemination, understanding the limitations of LLMs and the possibility of feeling surveilled by management. Unlike prior work that was forced to have managers mediate all interactions with operators~\cite{cheonWorkingBoundedCollaboration2022}, we were able to involve them separately in most cases. Thus, \textbf{we could maximize the openness and directness of the responses we received by positioning ourselves as external researchers (i.e., not `agents' of the management), building a relationship with participants over time, talking to operators without management present whenever possible, and maintaining the anonymity of participants}~\guideline{}.

From our work, we recognize that the most \textbf{crucial topic to involve operators is identifying what knowledge would be beneficial to share and how it can be expressed and retrieved}~\guideline{}. Tackling this together requires clear communication and education about what the technology can and cannot do. In this study, we witnessed the importance of respecting the operator's expertise and autonomy, making sure that operators feel their contributions through the CA are valued, not just as data for the system but as important knowledge that improves the workplace. For example, by focusing on sharing knowledge about resolving concrete machine issues as opposed to the less tangible machine tuning knowledge, more operators perceived the CA to be beneficial. This is especially important as many operators were apprehensive of introducing (yet) another tool and the potential consequences thereof, similar to observations made by~\cite{menewegerWorkingTogetherIndustrial2015}. By \textbf{ensuring everyone understands and sees the benefits of CAs, factories can reduce resistance and increase both productivity and job satisfaction}~\guideline{}.

Regarding social interactions, while AI allows for quick access to information, it could also lead to fewer face-to-face interactions, affecting the social environment. Additionally, increasing dependence on AI means operators must develop new skills, particularly in using advanced digital tools, instead of relying solely on experiential knowledge. This shift could also affect operators' well-being, balancing between the satisfaction from autonomously or collaboratively solving complex problems and the ease of AI help as previously discussed by~\citet{wurhoferReflectionsOperatorsMaintenance2018}. We observed that the operators were proud of their expertise, and introducing the CA to capture and share this might affect the satisfaction they get from work.

While using a CA for knowledge sharing promises to enhance operational efficiency, \textbf{it is vital to balance using the AI tool for all information needs versus the irreplaceable value of human knowledge sharing, if available}~\guideline{}. In fact, several operators preferred learning from human operators, similar to findings in prior work~\cite{hoffman2022retrofitt}. Yet, a key benefit of the tool is that it could help with problems when a knowledgeable human colleague is unavailable or unable to help.

The long-term effects of using AI for sharing knowledge in factories are likely complex, covering operational, psychological, and social aspects. Using AI for this purpose might improve operational efficiency and make the knowledge management system within organizations more dynamic and responsive. This could change how organizations are structured, making operators more independent of other (expert) operators. However, there may be fewer opportunities to share more subtle, tacit knowledge that is more challenging to share through technology, emphasizing the need for research in these areas.

Reliance on AI for decisions may also lead to poorer decision-making in the long-term, aligning with concerns raised by~\citet{buccinca2021totrust}. Furthermore, while an expert operator might intuitively be able to spot LLM `hallucinations', novice operators may not have sufficient experience to do so. Even if accurate, in the long term, overreliance on AI might weaken critical thinking and problem-solving abilities among operators. These challenges highlight the importance of longitudinal studies at factories. However, based on our experience navigating the safety and productivity concerns of factories during this study, this will be challenging to conduct. Aligning with prior work in this area, this highlights the need for methods that will allow the study of long-term impact in a way that is acceptable to industrial partners~\cite{cheonWorkingBoundedCollaboration2022}. 

\subsection{Limitations and Future Work}
In conducting this study, we faced numerous challenges tied to fieldwork in operational settings where financial, health, and safety concerns are paramount. Consequently, the tested prototypes were never utilized for extended periods during actual work activities; instead, we relied on a series of brief user tests, each lasting about 30 minutes. Despite the limitations posed by these short sessions, we are confident in the ecological validity of our findings given the high technology readiness levels (TRLs 7-8) of the prototypes, which were consistently integrated and tested within the real-world factory environment.

While our rapid prototyping approach facilitated immediate feedback and iterative improvements, a more prolonged deployment would likely uncover deeper insights into how users gradually adapt to and integrate CAs in their daily workflows. Such extended usage could provide a clearer picture of the shifts in user acceptance, the evolving effectiveness of the system, and the potential unforeseen impacts on workplace dynamics and operator autonomy.

Additionally, this study underlines the importance of socio-technical alignment when implementing AI-driven systems like CAs. In this study, we shifted toward sharing knowledge about problem-solving as all operators recognized it as important, and the benefits were clear. The operators were less enthusiastic about sharing knowledge about other areas, such as machine setup and tuning, due to the situatedness, temporal nature, and personal aspects of the knowledge. However, this knowledge is extremely valuable to factories and warrants focused research to explore the challenge of sharing this more complex knowledge. Furthermore, CSCW research should explore the long-term impacts of deploying such technologies, focusing on how they reshape organizational structures, job roles, and the balance between digital and human-driven problem-solving processes. This includes addressing the challenges of maintaining critical thinking and decision-making skills among workers in increasingly AI-supported environments and exploring strategies to balance technological assistance with essential human capabilities.

\section{Conclusion}
This longitudinal study highlights the complexities and potential of integrating Cognitive Assistants (CAs) in factories to support operators, emphasizing the necessity of addressing both technological efficacy and socio-technical dynamics for successful implementation. Efficient knowledge sharing through CAs can significantly expedite problem-solving and help novices become more autonomous, yet hinges on the systems' abilities and operator's perceptions. Key to the adoption of CAs is ensuring that surveillance concerns are addressed transparently, preferably in a setting where operators are comfortable sharing their perspectives. Involving all stakeholders, including worker's councils and operators, in the development of CAs is crucial for trust and acceptance, where stakeholders are sufficiently informed and educated to engage in discussions. Future research should explore the balance between AI-driven efficiency and the retention of valuable human interactions, alongside the long-term impacts of CAs on workplace dynamics and human decision-making abilities.

\begin{acks}

\end{acks}

\bibliographystyle{ACM-Reference-Format}
\bibliography{references}


\begin{thebibliography}{97}


\ifx \showCODEN    \undefined \def \showCODEN     #1{\unskip}     \fi
\ifx \showDOI      \undefined \def \showDOI       #1{#1}\fi
\ifx \showISBNx    \undefined \def \showISBNx     #1{\unskip}     \fi
\ifx \showISBNxiii \undefined \def \showISBNxiii  #1{\unskip}     \fi
\ifx \showISSN     \undefined \def \showISSN      #1{\unskip}     \fi
\ifx \showLCCN     \undefined \def \showLCCN      #1{\unskip}     \fi
\ifx \shownote     \undefined \def \shownote      #1{#1}          \fi
\ifx \showarticletitle \undefined \def \showarticletitle #1{#1}   \fi
\ifx \showURL      \undefined \def \showURL       {\relax}        \fi
\providecommand\bibfield[2]{#2}
\providecommand\bibinfo[2]{#2}
\providecommand\natexlab[1]{#1}
\providecommand\showeprint[2][]{arXiv:#2}

\bibitem[Abubakar et~al\mbox{.}(2019)]%
        {abubakarKnowledgeManagementDecisionmaking2019}
\bibfield{author}{\bibinfo{person}{Abubakar~Mohammed Abubakar}, \bibinfo{person}{Hamzah Elrehail}, \bibinfo{person}{Maher~Ahmad Alatailat}, {and} \bibinfo{person}{Alev Elçi}.} \bibinfo{year}{2019}\natexlab{}.
\newblock \showarticletitle{Knowledge Management, Decision-Making Style and Organizational Performance}.
\newblock  \bibinfo{volume}{4}, \bibinfo{number}{2} (\bibinfo{date}{04} \bibinfo{year}{2019}), \bibinfo{pages}{104--114}.
\newblock
\showISSN{2444-569X}
\urldef\tempurl%
\url{https://doi.org/10.1016/j.jik.2017.07.003}
\showDOI{\tempurl}


\bibitem[Ackerman et~al\mbox{.}(2013)]%
        {ackerman2013sharing}
\bibfield{author}{\bibinfo{person}{Mark~S Ackerman}, \bibinfo{person}{Juri Dachtera}, \bibinfo{person}{Volkmar Pipek}, {and} \bibinfo{person}{Volker Wulf}.} \bibinfo{year}{2013}\natexlab{}.
\newblock \showarticletitle{Sharing knowledge and expertise: The CSCW view of knowledge management}.
\newblock \bibinfo{journal}{\emph{Computer Supported Cooperative Work (CSCW)}}  \bibinfo{volume}{22} (\bibinfo{year}{2013}), \bibinfo{pages}{531--573}.
\newblock
\urldef\tempurl%
\url{https://doi.org/10.1007/s10606-013-9192-8}
\showDOI{\tempurl}


\bibitem[Alavi and Leidner(2001)]%
        {alaviReviewKnowledgeManagement2001}
\bibfield{author}{\bibinfo{person}{Maryam Alavi} {and} \bibinfo{person}{Dorothy~E. Leidner}.} \bibinfo{year}{2001}\natexlab{}.
\newblock \showarticletitle{Review: {{Knowledge Management}} and {{Knowledge Management Systems}}: {{Conceptual Foundations}} and {{Research Issues}}}.
\newblock  \bibinfo{volume}{25}, \bibinfo{number}{1} (\bibinfo{year}{2001}), \bibinfo{pages}{107--136}.
\newblock
\showISSN{0276-7783}
\urldef\tempurl%
\url{https://doi.org/10.2307/3250961}
\showDOI{\tempurl}
\showeprint[jstor]{3250961}


\bibitem[Angulo et~al\mbox{.}(2023)]%
        {anguloCognitive2023b}
\bibfield{author}{\bibinfo{person}{Cecilio Angulo}, \bibinfo{person}{Alejandro Chacón}, {and} \bibinfo{person}{Pere Ponsa}.} \bibinfo{year}{2023}\natexlab{}.
\newblock \showarticletitle{Towards a Cognitive Assistant Supporting Human Operators in the {{Artificial Intelligence}} of {{Things}}}.
\newblock   \bibinfo{volume}{21} (\bibinfo{date}{04} \bibinfo{year}{2023}), \bibinfo{pages}{100673}.
\newblock
\showISSN{2542-6605}
\urldef\tempurl%
\url{https://doi.org/10.1016/j.iot.2022.100673}
\showDOI{\tempurl}


\bibitem[Balayn et~al\mbox{.}(2022)]%
        {balayn2022ready}
\bibfield{author}{\bibinfo{person}{Agathe Balayn}, \bibinfo{person}{Gaole He}, \bibinfo{person}{Andrea Hu}, \bibinfo{person}{Jie Yang}, {and} \bibinfo{person}{Ujwal Gadiraju}.} \bibinfo{year}{2022}\natexlab{}.
\newblock \showarticletitle{Ready Player One! Eliciting Diverse Knowledge Using A Configurable Game}. In \bibinfo{booktitle}{\emph{Proceedings of the ACM Web Conference 2022}} \emph{(\bibinfo{series}{WWW '22})}. \bibinfo{publisher}{Association for Computing Machinery}, \bibinfo{address}{New York, NY, USA}, \bibinfo{pages}{1709–1719}.
\newblock
\showISBNx{9781450390965}
\urldef\tempurl%
\url{https://doi.org/10.1145/3485447.3512241}
\showDOI{\tempurl}


\bibitem[Baldauf et~al\mbox{.}(2021)]%
        {baldaufHumanInterventionsSmart2021}
\bibfield{author}{\bibinfo{person}{Matthias Baldauf}, \bibinfo{person}{Sebastian Müller}, \bibinfo{person}{Arne Seeliger}, \bibinfo{person}{Tobias Küng}, \bibinfo{person}{Andreas Michel}, {and} \bibinfo{person}{Werner Züllig}.} \bibinfo{year}{2021}\natexlab{}.
\newblock \showarticletitle{Human {{Interventions}} in the {{Smart Factory}} – {{A Case Study}} on {{Co-Designing Mobile}} and {{Wearable Monitoring Systems}} with {{Manufacturing Staff}}}. In \bibinfo{booktitle}{\emph{Extended {{Abstracts}} of the 2021 {{CHI Conference}} on {{Human Factors}} in {{Computing Systems}}}} (New York, NY, USA) \emph{(\bibinfo{series}{{{CHI EA}} '21})}. \bibinfo{publisher}{Association for Computing Machinery}, \bibinfo{pages}{1--6}.
\newblock
\showISBNx{978-1-4503-8095-9}
\urldef\tempurl%
\url{https://doi.org/10.1145/3411763.3451774}
\showDOI{\tempurl}


\bibitem[Bang et~al\mbox{.}(2023)]%
        {bang2023multitask}
\bibfield{author}{\bibinfo{person}{Yejin Bang}, \bibinfo{person}{Samuel Cahyawijaya}, \bibinfo{person}{Nayeon Lee}, \bibinfo{person}{Wenliang Dai}, \bibinfo{person}{Dan Su}, \bibinfo{person}{Bryan Wilie}, \bibinfo{person}{Holy Lovenia}, \bibinfo{person}{Ziwei Ji}, \bibinfo{person}{Tiezheng Yu}, \bibinfo{person}{Willy Chung}, \bibinfo{person}{Quyet~V. Do}, \bibinfo{person}{Yan Xu}, {and} \bibinfo{person}{Pascale Fung}.} \bibinfo{year}{2023}\natexlab{}.
\newblock \bibinfo{booktitle}{\emph{A Multitask, Multilingual, Multimodal Evaluation of ChatGPT on Reasoning, Hallucination, and Interactivity}}.
\newblock
\showeprint[arxiv]{2302.04023}~[cs.CL]


\bibitem[Becerra-Fernandez and Sabherwal(2014)]%
        {becerra2014knowledge}
\bibfield{author}{\bibinfo{person}{Irma Becerra-Fernandez} {and} \bibinfo{person}{Rajiv Sabherwal}.} \bibinfo{year}{2014}\natexlab{}.
\newblock \bibinfo{booktitle}{\emph{Knowledge management: Systems and processes}}.
\newblock \bibinfo{publisher}{Routledge}.
\newblock


\bibitem[Berger and von Garrel(2023)]%
        {bergerHowDesignValuebased2023}
\bibfield{author}{\bibinfo{person}{Patrick Berger} {and} \bibinfo{person}{Joerg von Garrel}.} \bibinfo{year}{2023}\natexlab{}.
\newblock \showarticletitle{How to Design a Value-Based {{Chatbot}} for the Manufacturing Industry: {{An}} Empirical Study of an Internal Assistance for Employees}.
\newblock  (\bibinfo{date}{11} \bibinfo{year}{2023}).
\newblock
\showISSN{0933-1875, 1610-1987}
\urldef\tempurl%
\url{https://doi.org/10.1007/s13218-023-00817-6}
\showDOI{\tempurl}


\bibitem[Bobrow and Whalen(2002)]%
        {bobrow2002community}
\bibfield{author}{\bibinfo{person}{Daniel~G Bobrow} {and} \bibinfo{person}{Jack Whalen}.} \bibinfo{year}{2002}\natexlab{}.
\newblock \showarticletitle{Community knowledge sharing in practice: the Eureka story}.
\newblock \bibinfo{journal}{\emph{Reflections: The SoL Journal}} \bibinfo{volume}{4}, \bibinfo{number}{2} (\bibinfo{year}{2002}), \bibinfo{pages}{47--59}.
\newblock


\bibitem[Boehner et~al\mbox{.}(2007)]%
        {boehner2007hci}
\bibfield{author}{\bibinfo{person}{Kirsten Boehner}, \bibinfo{person}{Janet Vertesi}, \bibinfo{person}{Phoebe Sengers}, {and} \bibinfo{person}{Paul Dourish}.} \bibinfo{year}{2007}\natexlab{}.
\newblock \showarticletitle{How HCI interprets the probes}. In \bibinfo{booktitle}{\emph{Proceedings of the SIGCHI conference on Human factors in computing systems}}. \bibinfo{pages}{1077--1086}.
\newblock
\urldef\tempurl%
\url{https://doi.org/10.1145/1240624.1240789}
\showDOI{\tempurl}


\bibitem[Brade\v{s}ko et~al\mbox{.}(2017)]%
        {curious.2017}
\bibfield{author}{\bibinfo{person}{Luka Brade\v{s}ko}, \bibinfo{person}{Michael Witbrock}, \bibinfo{person}{Janez Starc}, \bibinfo{person}{Zala Herga}, \bibinfo{person}{Marko Grobelnik}, {and} \bibinfo{person}{Dunja Mladeni\'{c}}.} \bibinfo{year}{2017}\natexlab{}.
\newblock \showarticletitle{Curious Cat--Mobile, Context-Aware Conversational Crowdsourcing Knowledge Acquisition}.
\newblock \bibinfo{journal}{\emph{ACM Trans. Inf. Syst.}} \bibinfo{volume}{35}, \bibinfo{number}{4}, Article \bibinfo{articleno}{33} (\bibinfo{date}{8} \bibinfo{year}{2017}), \bibinfo{numpages}{46}~pages.
\newblock
\showISSN{1046-8188}
\urldef\tempurl%
\url{https://doi.org/10.1145/3086686}
\showDOI{\tempurl}


\bibitem[Brahe and Schmidt(2007)]%
        {braheStoryWorkingWorkflow2007}
\bibfield{author}{\bibinfo{person}{Steen Brahe} {and} \bibinfo{person}{Kjeld Schmidt}.} \bibinfo{year}{2007}\natexlab{}.
\newblock \showarticletitle{The Story of a Working Workflow Management System}. In \bibinfo{booktitle}{\emph{Proceedings of the 2007 {{ACM International Conference}} on {{Supporting Group Work}}}} (New York, NY, USA) \emph{(\bibinfo{series}{{{GROUP}} '07})}. \bibinfo{publisher}{Association for Computing Machinery}, \bibinfo{pages}{249--258}.
\newblock
\showISBNx{978-1-59593-845-9}
\urldef\tempurl%
\url{https://doi.org/10.1145/1316624.1316661}
\showDOI{\tempurl}


\bibitem[Breque et~al\mbox{.}(2021)]%
        {brequeIndustrySustainableHumancentric2021}
\bibfield{author}{\bibinfo{person}{Maija Breque}, \bibinfo{person}{Lars de Nul}, {and} \bibinfo{person}{Athanasios Petridis}.} \bibinfo{year}{2021}\natexlab{}.
\newblock \bibinfo{booktitle}{\emph{Industry 5.0 - {{Towards}} a Sustainable, Human-Centric and Resilient {{European}} Industry - {{European Commission}}}}.
\newblock \bibinfo{type}{{T}echnical {R}eport}. \bibinfo{institution}{European Commission}.
\newblock


\bibitem[Brown and Duguid(2000)]%
        {brown2000balancing}
\bibfield{author}{\bibinfo{person}{John~Seely Brown} {and} \bibinfo{person}{Paul Duguid}.} \bibinfo{year}{2000}\natexlab{}.
\newblock \showarticletitle{Balancing act: How to capture knowledge without killing it}.
\newblock \bibinfo{journal}{\emph{Harvard business review}} \bibinfo{volume}{78}, \bibinfo{number}{3} (\bibinfo{year}{2000}), \bibinfo{pages}{73--80}.
\newblock


\bibitem[Brown et~al\mbox{.}(2020)]%
        {brown2020}
\bibfield{author}{\bibinfo{person}{Tom Brown}, \bibinfo{person}{Benjamin Mann}, \bibinfo{person}{Nick Ryder}, \bibinfo{person}{Melanie Subbiah}, \bibinfo{person}{Jared~D Kaplan}, \bibinfo{person}{Prafulla Dhariwal}, \bibinfo{person}{Arvind Neelakantan}, \bibinfo{person}{Pranav Shyam}, \bibinfo{person}{Girish Sastry}, \bibinfo{person}{Amanda Askell}, \bibinfo{person}{Sandhini Agarwal}, \bibinfo{person}{Ariel Herbert-Voss}, \bibinfo{person}{Gretchen Krueger}, \bibinfo{person}{Tom Henighan}, \bibinfo{person}{Rewon Child}, \bibinfo{person}{Aditya Ramesh}, \bibinfo{person}{Daniel Ziegler}, \bibinfo{person}{Jeffrey Wu}, \bibinfo{person}{Clemens Winter}, \bibinfo{person}{Chris Hesse}, \bibinfo{person}{Mark Chen}, \bibinfo{person}{Eric Sigler}, \bibinfo{person}{Mateusz Litwin}, \bibinfo{person}{Scott Gray}, \bibinfo{person}{Benjamin Chess}, \bibinfo{person}{Jack Clark}, \bibinfo{person}{Christopher Berner}, \bibinfo{person}{Sam McCandlish}, \bibinfo{person}{Alec Radford}, \bibinfo{person}{Ilya Sutskever}, {and}
  \bibinfo{person}{Dario Amodei}.} \bibinfo{year}{2020}\natexlab{}.
\newblock \showarticletitle{Language Models are Few-Shot Learners}. In \bibinfo{booktitle}{\emph{Advances in Neural Information Processing Systems}}, \bibfield{editor}{\bibinfo{person}{H.~Larochelle}, \bibinfo{person}{M.~Ranzato}, \bibinfo{person}{R.~Hadsell}, \bibinfo{person}{M.F. Balcan}, {and} \bibinfo{person}{H.~Lin}} (Eds.), Vol.~\bibinfo{volume}{33}. \bibinfo{publisher}{Curran Associates, Inc.}, \bibinfo{pages}{1877--1901}.
\newblock


\bibitem[Bu\c{c}inca et~al\mbox{.}(2021)]%
        {buccinca2021totrust}
\bibfield{author}{\bibinfo{person}{Zana Bu\c{c}inca}, \bibinfo{person}{Maja~Barbara Malaya}, {and} \bibinfo{person}{Krzysztof~Z. Gajos}.} \bibinfo{year}{2021}\natexlab{}.
\newblock \showarticletitle{To Trust or to Think: Cognitive Forcing Functions Can Reduce Overreliance on AI in AI-assisted Decision-making}.
\newblock \bibinfo{journal}{\emph{Proc. ACM Hum.-Comput. Interact.}} \bibinfo{volume}{5}, \bibinfo{number}{CSCW1}, Article \bibinfo{articleno}{188} (\bibinfo{date}{4} \bibinfo{year}{2021}), \bibinfo{numpages}{21}~pages.
\newblock
\urldef\tempurl%
\url{https://doi.org/10.1145/3449287}
\showDOI{\tempurl}


\bibitem[Cabitza et~al\mbox{.}(2021)]%
        {cabitza2021need}
\bibfield{author}{\bibinfo{person}{Federico Cabitza}, \bibinfo{person}{Andrea Campagner}, {and} \bibinfo{person}{Carla Simone}.} \bibinfo{year}{2021}\natexlab{}.
\newblock \showarticletitle{The need to move away from agential-AI: Empirical investigations, useful concepts and open issues}.
\newblock \bibinfo{journal}{\emph{International Journal of human-computer studies}}  \bibinfo{volume}{155} (\bibinfo{year}{2021}), \bibinfo{pages}{102696}.
\newblock
\urldef\tempurl%
\url{https://doi.org/10.1016/j.ijhcs.2021.102696}
\showDOI{\tempurl}


\bibitem[Casillo et~al\mbox{.}(2020)]%
        {casillo2020}
\bibfield{author}{\bibinfo{person}{Mario Casillo}, \bibinfo{person}{Francesco Colace}, \bibinfo{person}{Loretta Fabbri}, \bibinfo{person}{Marco Lombardi}, \bibinfo{person}{Alessandra Romano}, {and} \bibinfo{person}{Domenico Santaniello}.} \bibinfo{year}{2020}\natexlab{}.
\newblock \showarticletitle{Chatbot in Industry 4.0: An Approach for Training New Employees}. In \bibinfo{booktitle}{\emph{2020 IEEE International Conference on Teaching, Assessment, and Learning for Engineering (TALE)}}. \bibinfo{pages}{371--376}.
\newblock
\urldef\tempurl%
\url{https://doi.org/10.1109/TALE48869.2020.9368339}
\showDOI{\tempurl}


\bibitem[Castane et~al\mbox{.}(2023)]%
        {castaneASSISTANTProjectAI2023a}
\bibfield{author}{\bibinfo{person}{G. Castane}, \bibinfo{person}{A. Dolgui}, \bibinfo{person}{N. Kousi}, \bibinfo{person}{B. Meyers}, \bibinfo{person}{S. Thevenin}, \bibinfo{person}{E. Vyhmeister}, {and} \bibinfo{person}{P.-O. Ostberg}.} \bibinfo{year}{2023}\natexlab{}.
\newblock \showarticletitle{The {{ASSISTANT}} Project: {{AI}} for High Level Decisions in Manufacturing}.
\newblock  \bibinfo{volume}{61}, \bibinfo{number}{7} (\bibinfo{date}{04} \bibinfo{year}{2023}), \bibinfo{pages}{2288--2306}.
\newblock
\showISSN{0020-7543, 1366-588X}
\urldef\tempurl%
\url{https://doi.org/10.1080/00207543.2022.2069525}
\showDOI{\tempurl}


\bibitem[Chang et~al\mbox{.}(2012)]%
        {chang2012factors}
\bibfield{author}{\bibinfo{person}{Chun-Ming Chang}, \bibinfo{person}{Meng-Hsiang Hsu}, {and} \bibinfo{person}{Chia-Hui Yen}.} \bibinfo{year}{2012}\natexlab{}.
\newblock \showarticletitle{Factors affecting knowledge management success: the fit perspective}.
\newblock \bibinfo{journal}{\emph{Journal of Knowledge Management}} \bibinfo{volume}{16}, \bibinfo{number}{6} (\bibinfo{year}{2012}), \bibinfo{pages}{847--861}.
\newblock
\urldef\tempurl%
\url{https://doi.org/10.1108/13673271211276155}
\showDOI{\tempurl}


\bibitem[Cheon et~al\mbox{.}(2022)]%
        {cheonWorkingBoundedCollaboration2022}
\bibfield{author}{\bibinfo{person}{EunJeong Cheon}, \bibinfo{person}{Eike Schneiders}, {and} \bibinfo{person}{Mikael~B. Skov}.} \bibinfo{year}{2022}\natexlab{}.
\newblock \showarticletitle{Working with {{Bounded Collaboration}}: {{A Qualitative Study}} on {{How Collaboration}} Is {{Co-constructed}} around {{Collaborative Robots}} in {{Industry}}}.
\newblock   \bibinfo{volume}{6} (\bibinfo{date}{11} \bibinfo{year}{2022}), \bibinfo{pages}{369:1--369:34}.
\newblock
Issue CSCW2.
\urldef\tempurl%
\url{https://doi.org/10.1145/3555094}
\showDOI{\tempurl}


\bibitem[Colabianchi et~al\mbox{.}(2023)]%
        {colabianchiHumantechnologyIntegrationIndustrial2023}
\bibfield{author}{\bibinfo{person}{Silvia Colabianchi}, \bibinfo{person}{Andrea Tedeschi}, {and} \bibinfo{person}{Francesco Costantino}.} \bibinfo{year}{2023}\natexlab{}.
\newblock \showarticletitle{Human-Technology Integration with Industrial Conversational Agents: {{A}} Conceptual Architecture and a Taxonomy for Manufacturing}.
\newblock   \bibinfo{volume}{35} (\bibinfo{date}{10} \bibinfo{year}{2023}), \bibinfo{pages}{100510}.
\newblock
\showISSN{2452-414X}
\urldef\tempurl%
\url{https://doi.org/10.1016/j.jii.2023.100510}
\showDOI{\tempurl}


\bibitem[Cooke(1994)]%
        {cookeVarietiesKnowledgeElicitation1994}
\bibfield{author}{\bibinfo{person}{Nancy~J. Cooke}.} \bibinfo{year}{1994}\natexlab{}.
\newblock \showarticletitle{Varieties of Knowledge Elicitation Techniques}.
\newblock  \bibinfo{volume}{41}, \bibinfo{number}{6} (\bibinfo{date}{12} \bibinfo{year}{1994}), \bibinfo{pages}{801--849}.
\newblock
\showISSN{1071-5819}
\urldef\tempurl%
\url{https://doi.org/10.1006/ijhc.1994.1083}
\showDOI{\tempurl}


\bibitem[Crabtree et~al\mbox{.}(2017)]%
        {crabtree2017repacking}
\bibfield{author}{\bibinfo{person}{Andy Crabtree}, \bibinfo{person}{Peter Tolmie}, {and} \bibinfo{person}{Will Knight}.} \bibinfo{year}{2017}\natexlab{}.
\newblock \showarticletitle{Repacking ‘privacy’for a networked world}.
\newblock \bibinfo{journal}{\emph{Computer Supported Cooperative Work (CSCW)}}  \bibinfo{volume}{26} (\bibinfo{year}{2017}), \bibinfo{pages}{453--488}.
\newblock
\urldef\tempurl%
\url{https://doi.org/10.1007/s10606-017-9276-y}
\showDOI{\tempurl}


\bibitem[Cullen and Bryman(1988)]%
        {cullen1988knowledge}
\bibfield{author}{\bibinfo{person}{Jet Cullen} {and} \bibinfo{person}{Alan Bryman}.} \bibinfo{year}{1988}\natexlab{}.
\newblock \showarticletitle{The knowledge acquisition bottleneck: time for reassessment?}
\newblock \bibinfo{journal}{\emph{Expert Systems}} \bibinfo{volume}{5}, \bibinfo{number}{3} (\bibinfo{year}{1988}), \bibinfo{pages}{216--225}.
\newblock
\urldef\tempurl%
\url{https://doi.org/10.1111/j.1468-0394.1988.tb00065.x}
\showDOI{\tempurl}


\bibitem[Das~Swain et~al\mbox{.}(2024)]%
        {dasswainSensibleSensitiveAI2024}
\bibfield{author}{\bibinfo{person}{Vedant Das~Swain}, \bibinfo{person}{Lan Gao}, \bibinfo{person}{Abhirup Mondal}, \bibinfo{person}{Gregory~D. Abowd}, {and} \bibinfo{person}{Munmun De~Choudhury}.} \bibinfo{year}{2024}\natexlab{}.
\newblock \showarticletitle{Sensible and {{Sensitive AI}} for {{Worker Wellbeing}}: {{Factors}} That {{Inform Adoption}} and {{Resistance}} for {{Information Workers}}}. In \bibinfo{booktitle}{\emph{Proceedings of the {{CHI Conference}} on {{Human Factors}} in {{Computing Systems}}}} (New York, NY, USA) \emph{(\bibinfo{series}{{{CHI}} '24})}. \bibinfo{publisher}{Association for Computing Machinery}, \bibinfo{pages}{1--30}.
\newblock
\urldef\tempurl%
\url{https://doi.org/10.1145/3613904.3642716}
\showDOI{\tempurl}


\bibitem[Davis(1989)]%
        {davis1989perceived}
\bibfield{author}{\bibinfo{person}{Fred~D Davis}.} \bibinfo{year}{1989}\natexlab{}.
\newblock \showarticletitle{Perceived usefulness, perceived ease of use, and user acceptance of information technology}.
\newblock \bibinfo{journal}{\emph{MIS quarterly}} (\bibinfo{year}{1989}), \bibinfo{pages}{319--340}.
\newblock


\bibitem[Dhuieb et~al\mbox{.}(2013)]%
        {dhuiebDigitalFactoryAssistant2013}
\bibfield{author}{\bibinfo{person}{Mohamed~Anis Dhuieb}, \bibinfo{person}{Florent Laroche}, {and} \bibinfo{person}{Alain Bernard}.} \bibinfo{year}{2013}\natexlab{}.
\newblock \showarticletitle{Digital {{Factory Assistant}}: {{Conceptual Framework}} and {{Research Propositions}}}. In \bibinfo{booktitle}{\emph{{{Product Lifecycle Management for Society}} ({{PLM}} 2013)}} (Berlin, 2013), \bibfield{editor}{\bibinfo{person}{A.~Bernard}, \bibinfo{person}{L.~Rivest}, {and} \bibinfo{person}{D.~Dutta}} (Eds.), Vol.~\bibinfo{volume}{409}. \bibinfo{publisher}{Springer-Verlag Berlin}, \bibinfo{pages}{500--509}.
\newblock
\showISBNx{978-3-642-41501-2}
\showISSN{1868-4238, 1868-422X}


\bibitem[Duhigg(2016)]%
        {duhigg2016google}
\bibfield{author}{\bibinfo{person}{Charles Duhigg}.} \bibinfo{year}{2016}\natexlab{}.
\newblock \showarticletitle{What Google learned from its quest to build the perfect team}.
\newblock \bibinfo{journal}{\emph{The New York Times Magazine}} \bibinfo{volume}{26}, \bibinfo{number}{2016} (\bibinfo{year}{2016}), \bibinfo{pages}{2016}.
\newblock


\bibitem[Dyer and Nobeoka(2000)]%
        {dyer_creating_2000}
\bibfield{author}{\bibinfo{person}{Jeffrey~H. Dyer} {and} \bibinfo{person}{Kentaro Nobeoka}.} \bibinfo{year}{2000}\natexlab{}.
\newblock \showarticletitle{Creating and managing a high-performance knowledge-sharing network: the {Toyota} case}.
\newblock \bibinfo{journal}{\emph{Strategic Management Journal}} \bibinfo{volume}{21}, \bibinfo{number}{3} (\bibinfo{year}{2000}), \bibinfo{pages}{345--367}.
\newblock
\showISSN{1097-0266}
\urldef\tempurl%
\url{https://doi.org/10.1002/(SICI)1097-0266(200003)21:3<345::AID-SMJ96>3.0.CO;2-N}
\showDOI{\tempurl}


\bibitem[Edwards et~al\mbox{.}(2008)]%
        {edwards2008clustering}
\bibfield{author}{\bibinfo{person}{Brett Edwards}, \bibinfo{person}{Michael Zatorsky}, {and} \bibinfo{person}{Richi Nayak}.} \bibinfo{year}{2008}\natexlab{}.
\newblock \showarticletitle{Clustering and classification of maintenance logs using text data mining}.
\newblock \bibinfo{journal}{\emph{Volume 87-Data Mining and Analytics 2008}} (\bibinfo{year}{2008}), \bibinfo{pages}{193--199}.
\newblock
\showISBNx{9781920682682}


\bibitem[Edwards et~al\mbox{.}(2010)]%
        {edwards2010the}
\bibfield{author}{\bibinfo{person}{W.~Keith Edwards}, \bibinfo{person}{Mark~W. Newman}, {and} \bibinfo{person}{Erika~Shehan Poole}.} \bibinfo{year}{2010}\natexlab{}.
\newblock \showarticletitle{The Infrastructure Problem in HCI}. In \bibinfo{booktitle}{\emph{Proceedings of the SIGCHI Conference on Human Factors in Computing Systems}} (Atlanta, Georgia, USA) \emph{(\bibinfo{series}{CHI '10})}. \bibinfo{publisher}{Association for Computing Machinery}, \bibinfo{address}{New York, NY, USA}, \bibinfo{pages}{423–432}.
\newblock
\showISBNx{9781605589299}
\urldef\tempurl%
\url{https://doi.org/10.1145/1753326.1753390}
\showDOI{\tempurl}


\bibitem[Fenoglio et~al\mbox{.}(2022)]%
        {fenoglio_tacit_2022}
\bibfield{author}{\bibinfo{person}{Enzo Fenoglio}, \bibinfo{person}{Emre Kazim}, \bibinfo{person}{Hugo Latapie}, {and} \bibinfo{person}{Adriano Koshiyama}.} \bibinfo{year}{2022}\natexlab{}.
\newblock \showarticletitle{Tacit knowledge elicitation process for industry 4.0}.
\newblock \bibinfo{journal}{\emph{Discover Artificial Intelligence}} \bibinfo{volume}{2}, \bibinfo{number}{1} (\bibinfo{date}{3} \bibinfo{year}{2022}), \bibinfo{pages}{6}.
\newblock
\showISSN{2731-0809}
\urldef\tempurl%
\url{https://doi.org/10.1007/s44163-022-00020-w}
\showDOI{\tempurl}


\bibitem[Fereday and Muir-Cochrane(2006)]%
        {fereday2006demonstrating}
\bibfield{author}{\bibinfo{person}{Jennifer Fereday} {and} \bibinfo{person}{Eimear Muir-Cochrane}.} \bibinfo{year}{2006}\natexlab{}.
\newblock \showarticletitle{Demonstrating rigor using thematic analysis: A hybrid approach of inductive and deductive coding and theme development}.
\newblock \bibinfo{journal}{\emph{International journal of qualitative methods}} \bibinfo{volume}{5}, \bibinfo{number}{1} (\bibinfo{year}{2006}), \bibinfo{pages}{80--92}.
\newblock
\urldef\tempurl%
\url{https://doi.org/10.1177/160940690600500107}
\showDOI{\tempurl}


\bibitem[Fox et~al\mbox{.}(2020)]%
        {fox2020worker}
\bibfield{author}{\bibinfo{person}{Sarah~E. Fox}, \bibinfo{person}{Vera Khovanskaya}, \bibinfo{person}{Clara Crivellaro}, \bibinfo{person}{Niloufar Salehi}, \bibinfo{person}{Lynn Dombrowski}, \bibinfo{person}{Chinmay Kulkarni}, \bibinfo{person}{Lilly Irani}, {and} \bibinfo{person}{Jodi Forlizzi}.} \bibinfo{year}{2020}\natexlab{}.
\newblock \showarticletitle{Worker-Centered Design: Expanding HCI Methods for Supporting Labor} \emph{(\bibinfo{series}{CHI EA '20})}. \bibinfo{publisher}{Association for Computing Machinery}, \bibinfo{address}{New York, NY, USA}, \bibinfo{pages}{1–8}.
\newblock
\showISBNx{9781450368193}
\urldef\tempurl%
\url{https://doi.org/10.1145/3334480.3375157}
\showDOI{\tempurl}


\bibitem[Gao et~al\mbox{.}(2021)]%
        {gao-etal-2021-making}
\bibfield{author}{\bibinfo{person}{Tianyu Gao}, \bibinfo{person}{Adam Fisch}, {and} \bibinfo{person}{Danqi Chen}.} \bibinfo{year}{2021}\natexlab{}.
\newblock \showarticletitle{Making Pre-trained Language Models Better Few-shot Learners}. In \bibinfo{booktitle}{\emph{Proceedings of the 59th Annual Meeting of the Association for Computational Linguistics and the 11th International Joint Conference on Natural Language Processing (Volume 1: Long Papers)}}. \bibinfo{publisher}{Association for Computational Linguistics}, \bibinfo{address}{Online}, \bibinfo{pages}{3816--3830}.
\newblock
\urldef\tempurl%
\url{https://doi.org/10.18653/v1/2021.acl-long.295}
\showDOI{\tempurl}


\bibitem[Gerbracht et~al\mbox{.}(2024)]%
        {gerbrachtWeAreNo2024}
\bibfield{author}{\bibinfo{person}{Marc Gerbracht}, \bibinfo{person}{Max Krüger}, \bibinfo{person}{Débora De~Castro~Leal}, \bibinfo{person}{Peter Tolmie}, {and} \bibinfo{person}{Volker Wulf}.} \bibinfo{year}{2024}\natexlab{}.
\newblock \showarticletitle{"{{We}} Are No {{Luddites}}!" - {{CSCW}}, {{Co-Determination}} and {{Digital Transformation}} in {{Germany}}}.
\newblock   \bibinfo{volume}{8} (\bibinfo{date}{04} \bibinfo{year}{2024}), \bibinfo{pages}{153:1--153:23}.
\newblock
Issue CSCW1.
\urldef\tempurl%
\url{https://doi.org/10.1145/3637430}
\showDOI{\tempurl}


\bibitem[Hannola et~al\mbox{.}(2020)]%
        {hannolaAssessingImpactSociotechnical2020b}
\bibfield{author}{\bibinfo{person}{Lea Hannola}, \bibinfo{person}{Francisco Lacueva-Pérez}, \bibinfo{person}{Paolo Pretto}, \bibinfo{person}{Alexander Richter}, \bibinfo{person}{Marlene Schafler}, {and} \bibinfo{person}{Melanie Steinhüser}.} \bibinfo{year}{2020}\natexlab{}.
\newblock \showarticletitle{Assessing the Impact of Socio-Technical Interventions on Shop Floor Work Practices}.
\newblock  \bibinfo{volume}{33}, \bibinfo{number}{6} (\bibinfo{date}{06} \bibinfo{year}{2020}), \bibinfo{pages}{550--571}.
\newblock
\showISSN{0951-192X}
\urldef\tempurl%
\url{https://doi.org/10.1080/0951192X.2020.1775296}
\showDOI{\tempurl}


\bibitem[High(2012)]%
        {high2012era}
\bibfield{author}{\bibinfo{person}{Rob High}.} \bibinfo{year}{2012}\natexlab{}.
\newblock \showarticletitle{The era of cognitive systems: An inside look at IBM Watson and how it works}.
\newblock \bibinfo{journal}{\emph{IBM Corporation, Redbooks}}  \bibinfo{volume}{1} (\bibinfo{year}{2012}), \bibinfo{pages}{16}.
\newblock


\bibitem[Hoerner et~al\mbox{.}(2022)]%
        {hoerner_using_2022}
\bibfield{author}{\bibinfo{person}{Lorenz Hoerner}, \bibinfo{person}{Markus Schamberger}, {and} \bibinfo{person}{Freimut Bodendorf}.} \bibinfo{year}{2022}\natexlab{}.
\newblock \showarticletitle{Using {Tacit} {Expert} {Knowledge} to {Support} {Shop}-floor {Operators} {Through} a {Knowledge}-based {Assistance} {System}}.
\newblock \bibinfo{journal}{\emph{Computer Supported Cooperative Work (CSCW)}} (\bibinfo{date}{9} \bibinfo{year}{2022}).
\newblock
\showISSN{1573-7551}
\urldef\tempurl%
\url{https://doi.org/10.1007/s10606-022-09445-4}
\showDOI{\tempurl}


\bibitem[Hoffmann et~al\mbox{.}(2019)]%
        {hoffmannCyberPhysicalSystemsKnowledge2019a}
\bibfield{author}{\bibinfo{person}{Sven Hoffmann}, \bibinfo{person}{Aparecido Fabiano~Pinatti de Carvalho}, \bibinfo{person}{Darwin Abele}, \bibinfo{person}{Marcus Schweitzer}, \bibinfo{person}{Peter Tolmie}, {and} \bibinfo{person}{Volker Wulf}.} \bibinfo{year}{2019}\natexlab{}.
\newblock \showarticletitle{Cyber-{{Physical Systems}} for {{Knowledge}} and {{Expertise Sharing}} in {{Manufacturing Contexts}}: {{Towards}} a {{Model Enabling Design}}}.
\newblock  \bibinfo{volume}{28}, \bibinfo{number}{3-4} (\bibinfo{date}{06} \bibinfo{year}{2019}), \bibinfo{pages}{469--509}.
\newblock
\showISSN{0925-9724}
\urldef\tempurl%
\url{https://doi.org/10.1007/s10606-019-09355-y}
\showDOI{\tempurl}


\bibitem[Hoffmann et~al\mbox{.}(2022)]%
        {hoffman2022retrofitt}
\bibfield{author}{\bibinfo{person}{Sven Hoffmann}, \bibinfo{person}{Thomas Ludwig}, \bibinfo{person}{Florian Jasche}, \bibinfo{person}{Volker Wulf}, {and} \bibinfo{person}{David Randall}.} \bibinfo{year}{2022}\natexlab{}.
\newblock \showarticletitle{RetrofittAR: Supporting Hardware-Centered Expertise Sharing in Manufacturing Settings through Augmented Reality}.
\newblock \bibinfo{journal}{\emph{Comput. Supported Coop. Work}} \bibinfo{volume}{32}, \bibinfo{number}{1} (\bibinfo{date}{6} \bibinfo{year}{2022}), \bibinfo{pages}{93–139}.
\newblock
\showISSN{0925-9724}
\urldef\tempurl%
\url{https://doi.org/10.1007/s10606-022-09430-x}
\showDOI{\tempurl}


\bibitem[Jawahar et~al\mbox{.}(2019)]%
        {jawahar-etal-2019-bert}
\bibfield{author}{\bibinfo{person}{Ganesh Jawahar}, \bibinfo{person}{Beno{\^\i}t Sagot}, {and} \bibinfo{person}{Djam{\'e} Seddah}.} \bibinfo{year}{2019}\natexlab{}.
\newblock \showarticletitle{What Does {BERT} Learn about the Structure of Language?}. In \bibinfo{booktitle}{\emph{Proceedings of the 57th Annual Meeting of the Association for Computational Linguistics}}. \bibinfo{publisher}{Association for Computational Linguistics}, \bibinfo{address}{Florence, Italy}, \bibinfo{pages}{3651--3657}.
\newblock
\urldef\tempurl%
\url{https://doi.org/10.18653/v1/P19-1356}
\showDOI{\tempurl}


\bibitem[Kernan~Freire et~al\mbox{.}(2022)]%
        {KernanFreire2022ACognitive}
\bibfield{author}{\bibinfo{person}{Samuel Kernan~Freire}, \bibinfo{person}{Sarath~Surendranadha Panicker}, \bibinfo{person}{Santiago Ruiz-Arenas}, \bibinfo{person}{Zoltán Rusák}, {and} \bibinfo{person}{Evangelos Niforatos}.} \bibinfo{year}{2022}\natexlab{}.
\newblock \showarticletitle{A Cognitive Assistant for Operators: AI-Powered Knowledge Sharing on Complex Systems}.
\newblock \bibinfo{journal}{\emph{IEEE Pervasive Computing}} (\bibinfo{year}{2022}), \bibinfo{pages}{1--9}.
\newblock
\urldef\tempurl%
\url{https://doi.org/10.1109/MPRV.2022.3218600}
\showDOI{\tempurl}


\bibitem[Kiangala and Wang(2024)]%
        {kiangalaExperimentalHybridCustomized2024a}
\bibfield{author}{\bibinfo{person}{Kahiomba~Sonia Kiangala} {and} \bibinfo{person}{Zenghui Wang}.} \bibinfo{year}{2024}\natexlab{}.
\newblock \showarticletitle{An Experimental Hybrid Customized {{AI}} and Generative {{AI}} Chatbot Human Machine Interface to Improve a Factory Troubleshooting Downtime in the Context of {{Industry}} 5.0}.
\newblock  (\bibinfo{date}{04} \bibinfo{year}{2024}).
\newblock
\showISSN{0268-3768, 1433-3015}
\urldef\tempurl%
\url{https://doi.org/10.1007/s00170-024-13492-0}
\showDOI{\tempurl}


\bibitem[Kotthaus et~al\mbox{.}(2022)]%
        {kotthaus2022nego}
\bibfield{author}{\bibinfo{person}{Christoph Kotthaus}, \bibinfo{person}{Nico Vitt}, \bibinfo{person}{Max Kr\"{u}ger}, \bibinfo{person}{Volkmar Pipek}, {and} \bibinfo{person}{Volker Wulf}.} \bibinfo{year}{2022}\natexlab{}.
\newblock \showarticletitle{Negotiating Priorities on the Shopfloor: A Design Case Study of Maintainers’ Practices}.
\newblock \bibinfo{journal}{\emph{Comput. Supported Coop. Work}} \bibinfo{volume}{32}, \bibinfo{number}{1} (\bibinfo{date}{10} \bibinfo{year}{2022}), \bibinfo{pages}{141–210}.
\newblock
\showISSN{0925-9724}
\urldef\tempurl%
\url{https://doi.org/10.1007/s10606-022-09444-5}
\showDOI{\tempurl}


\bibitem[Lakhani et~al\mbox{.}(2013)]%
        {lakhani2013open}
\bibfield{author}{\bibinfo{person}{Karim~R Lakhani}, \bibinfo{person}{Katja Hutter}, \bibinfo{person}{S~Healy Pokrywa}, {and} \bibinfo{person}{Johann Fuller}.} \bibinfo{year}{2013}\natexlab{}.
\newblock \showarticletitle{Open innovation at Siemens}.
\newblock \bibinfo{journal}{\emph{Harvard Business School Case}}  \bibinfo{volume}{613} (\bibinfo{year}{2013}), \bibinfo{pages}{100}.
\newblock


\bibitem[Lewis et~al\mbox{.}(2020)]%
        {lewis2020retr}
\bibfield{author}{\bibinfo{person}{Patrick Lewis}, \bibinfo{person}{Ethan Perez}, \bibinfo{person}{Aleksandra Piktus}, \bibinfo{person}{Fabio Petroni}, \bibinfo{person}{Vladimir Karpukhin}, \bibinfo{person}{Naman Goyal}, \bibinfo{person}{Heinrich K\"{u}ttler}, \bibinfo{person}{Mike Lewis}, \bibinfo{person}{Wen-tau Yih}, \bibinfo{person}{Tim Rockt\"{a}schel}, \bibinfo{person}{Sebastian Riedel}, {and} \bibinfo{person}{Douwe Kiela}.} \bibinfo{year}{2020}\natexlab{}.
\newblock \showarticletitle{Retrieval-Augmented Generation for Knowledge-Intensive NLP Tasks}. In \bibinfo{booktitle}{\emph{Proceedings of the 34th International Conference on Neural Information Processing Systems}} (Vancouver, BC, Canada) \emph{(\bibinfo{series}{NIPS'20})}. \bibinfo{publisher}{Curran Associates Inc.}, \bibinfo{address}{Red Hook, NY, USA}, Article \bibinfo{articleno}{793}, \bibinfo{numpages}{16}~pages.
\newblock
\showISBNx{9781713829546}


\bibitem[Leyer et~al\mbox{.}(2018)]%
        {leyerPowerWorkersEmpowering2018}
\bibfield{author}{\bibinfo{person}{Michael Leyer}, \bibinfo{person}{Alexander Richter}, {and} \bibinfo{person}{Melanie Steinhüser}.} \bibinfo{year}{2018}\natexlab{}.
\newblock \showarticletitle{“{{Power}} to the Workers”: {{Empowering}} Shop Floor Workers with Worker-Centric Digital Designs}.
\newblock  \bibinfo{volume}{39}, \bibinfo{number}{1} (\bibinfo{date}{01} \bibinfo{year}{2018}), \bibinfo{pages}{24--42}.
\newblock
\showISSN{0144-3577}
\urldef\tempurl%
\url{https://doi.org/10.1108/IJOPM-05-2017-0294}
\showDOI{\tempurl}


\bibitem[Li et~al\mbox{.}(2022)]%
        {liBringingNaturalLanguageenabled2022}
\bibfield{author}{\bibinfo{person}{Chen Li}, \bibinfo{person}{Andreas~Kornmaaler Hansen}, \bibinfo{person}{Dimitrios Chrysostomou}, \bibinfo{person}{Simon Bogh}, {and} \bibinfo{person}{Ole Madsen}.} \bibinfo{year}{2022}\natexlab{}.
\newblock \showarticletitle{Bringing a {{Natural Language-enabled Virtual Assistant}} to {{Industrial Mobile Robots}} for {{Learning}}, {{Training}} and {{Assistance}} of {{Manufacturing Tasks}}}. In \bibinfo{booktitle}{\emph{2022 {{IEEE}}/{{SICE International Symposium on System Integration}} ({{SII}} 2022)}} (New York, 2022). \bibinfo{publisher}{IEEE}, \bibinfo{pages}{238--243}.
\newblock
\showISBNx{978-1-66544-540-5}
\showISSN{2474-2317}
\urldef\tempurl%
\url{https://doi.org/10.1109/SII52469.2022.9708757}
\showDOI{\tempurl}


\bibitem[Li and Yang(2021)]%
        {liBotXAIbasedVirtual2021a}
\bibfield{author}{\bibinfo{person}{Chen Li} {and} \bibinfo{person}{Hong~Ji Yang}.} \bibinfo{year}{2021}\natexlab{}.
\newblock \showarticletitle{Bot-{{X}}: {{An AI-based}} Virtual Assistant for Intelligent Manufacturing}.
\newblock  \bibinfo{volume}{17}, \bibinfo{number}{1} (\bibinfo{year}{2021}), \bibinfo{pages}{1--14}.
\newblock
\showISSN{1574-1702, 1875-9076}
\urldef\tempurl%
\url{https://doi.org/10.3233/MGS-210340}
\showDOI{\tempurl}


\bibitem[Liebowitz(2002)]%
        {liebowitz2002nasa}
\bibfield{author}{\bibinfo{person}{J. Liebowitz}.} \bibinfo{year}{2002}\natexlab{}.
\newblock \showarticletitle{A look at NASA Goddard Space Flight Center's knowledge management initiatives}.
\newblock \bibinfo{journal}{\emph{IEEE Software}} \bibinfo{volume}{19}, \bibinfo{number}{3} (\bibinfo{year}{2002}), \bibinfo{pages}{40--42}.
\newblock
\urldef\tempurl%
\url{https://doi.org/10.1109/MS.2002.1003451}
\showDOI{\tempurl}


\bibitem[Lindsay et~al\mbox{.}(1993)]%
        {LINDSAY1993209}
\bibfield{author}{\bibinfo{person}{Robert~K. Lindsay}, \bibinfo{person}{Bruce~G. Buchanan}, \bibinfo{person}{Edward~A. Feigenbaum}, {and} \bibinfo{person}{Joshua Lederberg}.} \bibinfo{year}{1993}\natexlab{}.
\newblock \showarticletitle{DENDRAL: A case study of the first expert system for scientific hypothesis formation}.
\newblock \bibinfo{journal}{\emph{Artificial Intelligence}} \bibinfo{volume}{61}, \bibinfo{number}{2} (\bibinfo{year}{1993}), \bibinfo{pages}{209--261}.
\newblock
\showISSN{0004-3702}
\urldef\tempurl%
\url{https://doi.org/10.1016/0004-3702(93)90068-M}
\showDOI{\tempurl}


\bibitem[Lista~Rossetti et~al\mbox{.}(2023)]%
        {listarossettiIdentifyingIndustryTechnologies2023}
\bibfield{author}{\bibinfo{person}{Ana~Paula Lista~Rossetti}, \bibinfo{person}{Guilherme Luz~Tortorella}, \bibinfo{person}{Marina Bouzon}, \bibinfo{person}{Shang Gao}, {and} \bibinfo{person}{Toong~Khuan Chan}.} \bibinfo{year}{2023}\natexlab{}.
\newblock \showarticletitle{Identifying {{Industry}} 4.0 Technologies Enablers for\textasciitilde Knowledge Management –\textasciitilde a\textasciitilde Scoping Review}.
\newblock  \bibinfo{volume}{36}, \bibinfo{number}{1} (\bibinfo{date}{01} \bibinfo{year}{2023}), \bibinfo{pages}{340--360}.
\newblock
\showISSN{1754-2731}
\urldef\tempurl%
\url{https://doi.org/10.1108/TQM-05-2022-0173}
\showDOI{\tempurl}


\bibitem[Ludwig et~al\mbox{.}(2017)]%
        {ludwig20173D}
\bibfield{author}{\bibinfo{person}{Thomas Ludwig}, \bibinfo{person}{Alexander Boden}, {and} \bibinfo{person}{Volkmar Pipek}.} \bibinfo{year}{2017}\natexlab{}.
\newblock \showarticletitle{3D Printers as Sociable Technologies: Taking Appropriation Infrastructures to the Internet of Things}.
\newblock  \bibinfo{volume}{24}, \bibinfo{number}{2}, Article \bibinfo{articleno}{17} (\bibinfo{date}{4} \bibinfo{year}{2017}), \bibinfo{numpages}{28}~pages.
\newblock
\showISSN{1073-0516}
\urldef\tempurl%
\url{https://doi.org/10.1145/3007205}
\showDOI{\tempurl}


\bibitem[Maedche et~al\mbox{.}(2019)]%
        {maedche2019ai}
\bibfield{author}{\bibinfo{person}{Alexander Maedche}, \bibinfo{person}{Christine Legner}, \bibinfo{person}{Alexander Benlian}, \bibinfo{person}{Benedikt Berger}, \bibinfo{person}{Henner Gimpel}, \bibinfo{person}{Thomas Hess}, \bibinfo{person}{Oliver Hinz}, \bibinfo{person}{Stefan Morana}, {and} \bibinfo{person}{Matthias S{\"o}llner}.} \bibinfo{year}{2019}\natexlab{}.
\newblock \showarticletitle{AI-based digital assistants: Opportunities, threats, and research perspectives}.
\newblock \bibinfo{journal}{\emph{Business \& Information Systems Engineering}}  \bibinfo{volume}{61} (\bibinfo{year}{2019}), \bibinfo{pages}{535--544}.
\newblock
\urldef\tempurl%
\url{https://doi.org/10.1007/s12599-019-00600-8}
\showDOI{\tempurl}


\bibitem[Mantravadi et~al\mbox{.}(2020)]%
        {mantravadiUserFriendlyMESInterfaces2020j}
\bibfield{author}{\bibinfo{person}{Soujanya Mantravadi}, \bibinfo{person}{Andreas~Dyroy Jansson}, {and} \bibinfo{person}{Charles Moller}.} \bibinfo{year}{2020}\natexlab{}.
\newblock \showarticletitle{User-{{Friendly MES Interfaces}}: {{Recommendations}} for an {{AI-Based Chatbot Assistance}} in {{Industry}} 4.0 {{Shop Floors}}}. In \bibinfo{booktitle}{\emph{{{Intelligent Information and Database Systems}} ({{ACIIDS}} 2020), {{PT II}}}} (Cham, 2020), \bibfield{editor}{\bibinfo{person}{N.~T. Nguyen}, \bibinfo{person}{K.~Jearanaitanakij}, \bibinfo{person}{A.~Selamat}, \bibinfo{person}{B.~Trawinski}, {and} \bibinfo{person}{S.~Chittayasothorn}} (Eds.), Vol.~\bibinfo{volume}{12034}. \bibinfo{publisher}{Springer International Publishing Ag}, \bibinfo{pages}{189--201}.
\newblock
\showISSN{0302-9743, 1611-3349}
\urldef\tempurl%
\url{https://doi.org/10.1007/978-3-030-42058-1\_16}
\showDOI{\tempurl}


\bibitem[Martinez-Maldonado et~al\mbox{.}(2023)]%
        {martinez-maldonado_lessons_2023}
\bibfield{author}{\bibinfo{person}{Roberto Martinez-Maldonado}, \bibinfo{person}{Vanessa Echeverria}, \bibinfo{person}{Gloria Fernandez-Nieto}, \bibinfo{person}{Lixiang Yan}, \bibinfo{person}{Linxuan Zhao}, \bibinfo{person}{Riordan Alfredo}, \bibinfo{person}{Xinyu Li}, \bibinfo{person}{Samantha Dix}, \bibinfo{person}{Hollie Jaggard}, \bibinfo{person}{Rosie Wotherspoon}, \bibinfo{person}{Abra Osborne}, \bibinfo{person}{Simon~Buckingham Shum}, {and} \bibinfo{person}{Dragan Gašević}.} \bibinfo{year}{2023}\natexlab{}.
\newblock \showarticletitle{Lessons {Learnt} from a {Multimodal} {Learning} {Analytics} {Deployment} {In}-the-wild}.
\newblock \bibinfo{journal}{\emph{ACM Transactions on Computer-Human Interaction}} (\bibinfo{date}{9} \bibinfo{year}{2023}).
\newblock
\showISSN{1073-0516}
\urldef\tempurl%
\url{https://doi.org/10.1145/3622784}
\showDOI{\tempurl}
\newblock
\shownote{Just Accepted}.


\bibitem[Massey et~al\mbox{.}(2001)]%
        {masseyReengineeringCustomerRelationship2001}
\bibfield{author}{\bibinfo{person}{Anne~P. Massey}, \bibinfo{person}{Mitzi~M. Montoya-Weiss}, {and} \bibinfo{person}{Kent Holcom}.} \bibinfo{year}{2001}\natexlab{}.
\newblock \showarticletitle{Re-Engineering the Customer Relationship: Leveraging Knowledge Assets at {{IBM}}}.
\newblock  \bibinfo{volume}{32}, \bibinfo{number}{2} (\bibinfo{date}{12} \bibinfo{year}{2001}), \bibinfo{pages}{155--170}.
\newblock
\urldef\tempurl%
\url{https://doi.org/10.1016/S0167-9236(01)00108-7}
\showDOI{\tempurl}


\bibitem[Meneweger et~al\mbox{.}(2015)]%
        {menewegerWorkingTogetherIndustrial2015}
\bibfield{author}{\bibinfo{person}{Thomas Meneweger}, \bibinfo{person}{Daniela Wurhofer}, \bibinfo{person}{Verena Fuchsberger}, {and} \bibinfo{person}{Manfred Tscheligi}.} \bibinfo{year}{2015}\natexlab{}.
\newblock \showarticletitle{Working Together with Industrial Robots: {{Experiencing}} Robots in a Production Environment}. In \bibinfo{booktitle}{\emph{2015 24th {{IEEE International Symposium}} on {{Robot}} and {{Human Interactive Communication}} ({{RO-MAN}})}}. \bibinfo{pages}{833--838}.
\newblock
\urldef\tempurl%
\url{https://doi.org/10.1109/ROMAN.2015.7333641}
\showDOI{\tempurl}


\bibitem[Mercedes-Benz(2023)]%
        {mercedes2023}
\bibfield{author}{\bibinfo{person}{Mercedes-Benz}.} \bibinfo{year}{2023}\natexlab{}.
\newblock \bibinfo{title}{Benz tests chatgpt in intelligent vehicle production.: Mercedes-Benz Group}.
\newblock
\newblock


\bibitem[Mucha et~al\mbox{.}(2018)]%
        {muchaIndustrialInternetThings2018}
\bibfield{author}{\bibinfo{person}{Henrik Mucha}, \bibinfo{person}{Carsten Roecker}, \bibinfo{person}{Thomas Ludwig}, \bibinfo{person}{Corinna Ogonowski}, \bibinfo{person}{Martin Stein}, \bibinfo{person}{Sebastian Robert}, \bibinfo{person}{Lukas Galla}, \bibinfo{person}{Martin Hill}, {and} \bibinfo{person}{Volker Wulf}.} \bibinfo{year}{2018}\natexlab{}.
\newblock \showarticletitle{The {{Industrial Internet}} of {{Things}}: {{New Perspectives}} on {{HCI}} and {{CSCW}} within {{Industry Settings}}}. In \bibinfo{booktitle}{\emph{Companion of the 2018 {{ACM Conference}} on {{Computer Supported Cooperative Work}} and {{Social Computing}}}} (New York, NY, USA) \emph{(\bibinfo{series}{{{CSCW}} '18 {{Companion}}})}. \bibinfo{publisher}{Association for Computing Machinery}, \bibinfo{pages}{393--400}.
\newblock
\showISBNx{978-1-4503-6018-0}
\urldef\tempurl%
\url{https://doi.org/10.1145/3272973.3273009}
\showDOI{\tempurl}


\bibitem[Mörike(2022)]%
        {morikeInvertedHierarchiesShop2022}
\bibfield{author}{\bibinfo{person}{Frauke Mörike}.} \bibinfo{year}{2022}\natexlab{}.
\newblock \showarticletitle{Inverted {{Hierarchies}} on the {{Shop Floor}}: {{The Organisational Layer}} of {{Workarounds}} for {{Collaboration}} in the {{Metal Industry}}}.
\newblock  \bibinfo{volume}{31}, \bibinfo{number}{1} (\bibinfo{date}{03} \bibinfo{year}{2022}), \bibinfo{pages}{111--147}.
\newblock
\showISSN{0925-9724}
\urldef\tempurl%
\url{https://doi.org/10.1007/s10606-021-09415-2}
\showDOI{\tempurl}


\bibitem[Müller et~al\mbox{.}(2021)]%
        {mullerDigitalShopFloor2021}
\bibfield{author}{\bibinfo{person}{Marvin Müller}, \bibinfo{person}{Emanuel Alexandi}, {and} \bibinfo{person}{Joachim Metternich}.} \bibinfo{year}{2021}\natexlab{}.
\newblock \showarticletitle{Digital Shop Floor Management Enhanced by Natural Language Processing}.
\newblock   \bibinfo{volume}{96} (\bibinfo{date}{01} \bibinfo{year}{2021}), \bibinfo{pages}{21--26}.
\newblock
\showISSN{2212-8271}
\urldef\tempurl%
\url{https://doi.org/10.1016/j.procir.2021.01.046}
\showDOI{\tempurl}


\bibitem[Naqvi et~al\mbox{.}(2022)]%
        {naqviHumanKnowledgeCentered2022a}
\bibfield{author}{\bibinfo{person}{Syed Meesam~Raza Naqvi}, \bibinfo{person}{Mohammad Ghufran}, \bibinfo{person}{Safa Meraghni}, \bibinfo{person}{Christophe Varnier}, \bibinfo{person}{Jean-Marc Nicod}, {and} \bibinfo{person}{Noureddine Zerhouni}.} \bibinfo{year}{2022}\natexlab{}.
\newblock \showarticletitle{Human Knowledge Centered Maintenance Decision Support in Digital Twin Environment}.
\newblock   \bibinfo{volume}{65} (\bibinfo{date}{10} \bibinfo{year}{2022}), \bibinfo{pages}{528--537}.
\newblock
\showISSN{0278-6125, 1878-6642}
\urldef\tempurl%
\url{https://doi.org/10.1016/j.jmsy.2022.10.003}
\showDOI{\tempurl}


\bibitem[Nonaka(2009)]%
        {nonaka2009knowledge}
\bibfield{author}{\bibinfo{person}{Ikujiro Nonaka}.} \bibinfo{year}{2009}\natexlab{}.
\newblock \showarticletitle{The knowledge-creating company}.
\newblock In \bibinfo{booktitle}{\emph{The economic impact of knowledge}}. \bibinfo{publisher}{Routledge}, \bibinfo{pages}{175--187}.
\newblock


\bibitem[Nonaka and Takeuchi(1995)]%
        {Nonaka.1995}
\bibfield{author}{\bibinfo{person}{Ikujiro Nonaka} {and} \bibinfo{person}{Hirotaka Takeuchi}.} \bibinfo{year}{1995}\natexlab{}.
\newblock \showarticletitle{The knowledge-creating company: How Japanese companies create the dynamics of innovation}.
\newblock \bibinfo{journal}{\emph{New York, NY}} (\bibinfo{year}{1995}).
\newblock


\bibitem[OpenAI(2023)]%
        {openai2023gpt4}
\bibfield{author}{\bibinfo{person}{OpenAI}.} \bibinfo{year}{2023}\natexlab{}.
\newblock \bibinfo{booktitle}{\emph{GPT-4 Technical Report}}.
\newblock
\showeprint[arxiv]{2303.08774}~[cs.CL]


\bibitem[Oru{\c{c}}(2020)]%
        {orucc2020semantic}
\bibfield{author}{\bibinfo{person}{Or{\c{c}}un Oru{\c{c}}}.} \bibinfo{year}{2020}\natexlab{}.
\newblock \showarticletitle{A semantic Question Answering through Heterogeneous data source in the domain of smart factory}.
\newblock \bibinfo{journal}{\emph{International Journal on Natural Language Computing (IJNLC) Vol}}  \bibinfo{volume}{9} (\bibinfo{year}{2020}).
\newblock


\bibitem[Pavlov et~al\mbox{.}(2020)]%
        {pavlovCaseStudyUsing2020}
\bibfield{author}{\bibinfo{person}{Dmitry Pavlov}, \bibinfo{person}{Igor Sosnovsky}, \bibinfo{person}{Vyacheslav Dimitrov}, \bibinfo{person}{Vasilii Melentyev}, {and} \bibinfo{person}{Dmitry Korzun}.} \bibinfo{year}{2020}\natexlab{}.
\newblock \showarticletitle{Case {{Study}} of {{Using Virtual}} and {{Augmented Reality}} in {{Industrial System Monitoring}}}. In \bibinfo{booktitle}{\emph{2020 26th {{Conference}} of {{Open Innovations Association}} ({{FRUCT}})}}. \bibinfo{pages}{367--375}.
\newblock
\showISSN{2305-7254}
\urldef\tempurl%
\url{https://doi.org/10.23919/FRUCT48808.2020.9087410}
\showDOI{\tempurl}


\bibitem[Probst et~al\mbox{.}(2000)]%
        {probstManagingKnowledgeBuilding2000}
\bibfield{author}{\bibinfo{person}{Gilbert J.~B. Probst}, \bibinfo{person}{Steffen Raub}, {and} \bibinfo{person}{Kai Romhardt}.} \bibinfo{year}{2000}\natexlab{}.
\newblock \bibinfo{booktitle}{\emph{Managing {{Knowledge}}: {{Building Blocks}} for {{Success}}}}.
\newblock \bibinfo{publisher}{Wiley}.
\newblock
\showISBNx{978-0-471-99768-9}
\showeprint[googlebooks]{aXyFQgAACAAJ}


\bibitem[Ras et~al\mbox{.}(2017)]%
        {rasBridgingSkillsGap2017}
\bibfield{author}{\bibinfo{person}{Eric Ras}, \bibinfo{person}{Fridolin Wild}, \bibinfo{person}{Christoph Stahl}, {and} \bibinfo{person}{Alexandre Baudet}.} \bibinfo{year}{2017}\natexlab{}.
\newblock \showarticletitle{Bridging the {{Skills Gap}} of {{Workers}} in {{Industry}} 4.0 by {{Human Performance Augmentation Tools}}: {{Challenges}} and {{Roadmap}}}. In \bibinfo{booktitle}{\emph{Proceedings of the 10th {{International Conference}} on {{Pervasive Technologies Related}} to {{Assistive Environments}}}} (New York, NY, USA) \emph{(\bibinfo{series}{{{PETRA}} '17})}. \bibinfo{publisher}{Association for Computing Machinery}, \bibinfo{pages}{428--432}.
\newblock
\showISBNx{978-1-4503-5227-7}
\urldef\tempurl%
\url{https://doi.org/10.1145/3056540.3076192}
\showDOI{\tempurl}


\bibitem[Reis et~al\mbox{.}(2022)]%
        {reisVirtualAssistanceContext2022}
\bibfield{author}{\bibinfo{person}{Arsénio Reis}, \bibinfo{person}{João Barroso}, \bibinfo{person}{Arlindo Santos}, \bibinfo{person}{Paulo Rodrigues}, {and} \bibinfo{person}{Rodrigo Pereira}.} \bibinfo{year}{2022}\natexlab{}.
\newblock \showarticletitle{Virtual {{Assistance}} in the {{Context}} of the {{Industry}} 4.0: {{A Case Study}} at {{Continental Advanced Antenna}}}. In \bibinfo{booktitle}{\emph{Information {{Systems}} and {{Technologies}}}} (Cham, 2022), \bibfield{editor}{\bibinfo{person}{Alvaro Rocha}, \bibinfo{person}{Hojjat Adeli}, \bibinfo{person}{Gintautas Dzemyda}, {and} \bibinfo{person}{Fernando Moreira}} (Eds.). \bibinfo{publisher}{Springer International Publishing}, \bibinfo{pages}{651--662}.
\newblock
\showISBNx{978-3-031-04826-5}
\urldef\tempurl%
\url{https://doi.org/10.1007/978-3-031-04826-5\_64}
\showDOI{\tempurl}


\bibitem[Rooein et~al\mbox{.}(2020)]%
        {rooeinChattingProcessesDigital2020}
\bibfield{author}{\bibinfo{person}{Donya Rooein}, \bibinfo{person}{Devis Bianchini}, \bibinfo{person}{Francesco Leotta}, \bibinfo{person}{Massimo Mecella}, \bibinfo{person}{Paolo Paolini}, {and} \bibinfo{person}{Barbara Pernici}.} \bibinfo{year}{2020}\natexlab{}.
\newblock \showarticletitle{Chatting {{About Processes}} in {{Digital Factories}}: {{A Model-Based Approach}}}. In \bibinfo{booktitle}{\emph{Enterprise, {{Business-Process}} and {{Information Systems Modeling}}}} (Cham, 2020), \bibfield{editor}{\bibinfo{person}{Selmin Nurcan}, \bibinfo{person}{Iris Reinhartz-Berger}, \bibinfo{person}{Pnina Soffer}, {and} \bibinfo{person}{Jelena Zdravkovic}} (Eds.). \bibinfo{publisher}{Springer International Publishing}, \bibinfo{pages}{70--84}.
\newblock
\showISBNx{978-3-030-49418-6}
\urldef\tempurl%
\url{https://doi.org/10.1007/978-3-030-49418-6\_5}
\showDOI{\tempurl}


\bibitem[Rosso et~al\mbox{.}(2010)]%
        {rossoMeaningWorkTheoretical2010}
\bibfield{author}{\bibinfo{person}{Brent~D. Rosso}, \bibinfo{person}{Kathryn~H. Dekas}, {and} \bibinfo{person}{Amy Wrzesniewski}.} \bibinfo{year}{2010}\natexlab{}.
\newblock \showarticletitle{On the Meaning of Work: {{A}} Theoretical Integration and Review}.
\newblock   \bibinfo{volume}{30} (\bibinfo{date}{01} \bibinfo{year}{2010}), \bibinfo{pages}{91--127}.
\newblock
\showISSN{0191-3085}
\urldef\tempurl%
\url{https://doi.org/10.1016/j.riob.2010.09.001}
\showDOI{\tempurl}


\bibitem[Schniederjans et~al\mbox{.}(2020)]%
        {schniederjansSupplyChainDigitisation2020}
\bibfield{author}{\bibinfo{person}{Dara~G. Schniederjans}, \bibinfo{person}{Carla Curado}, {and} \bibinfo{person}{Mehrnaz Khalajhedayati}.} \bibinfo{year}{2020}\natexlab{}.
\newblock \showarticletitle{Supply Chain Digitisation Trends: {{An}} Integration of Knowledge Management}.
\newblock   \bibinfo{volume}{220} (\bibinfo{date}{02} \bibinfo{year}{2020}), \bibinfo{pages}{107439}.
\newblock
\showISSN{0925-5273}
\urldef\tempurl%
\url{https://doi.org/10.1016/j.ijpe.2019.07.012}
\showDOI{\tempurl}


\bibitem[Serrat(2017)]%
        {serrat2017five}
\bibfield{author}{\bibinfo{person}{Olivier Serrat}.} \bibinfo{year}{2017}\natexlab{}.
\newblock \showarticletitle{The five whys technique}.
\newblock \bibinfo{journal}{\emph{Knowledge solutions: Tools, methods, and approaches to drive organizational performance}} (\bibinfo{year}{2017}), \bibinfo{pages}{307--310}.
\newblock


\bibitem[Sexton et~al\mbox{.}(2017)]%
        {8258120}
\bibfield{author}{\bibinfo{person}{Thurston Sexton}, \bibinfo{person}{Michael~P. Brundage}, \bibinfo{person}{Michael Hoffman}, {and} \bibinfo{person}{K~C Morris}.} \bibinfo{year}{2017}\natexlab{}.
\newblock \showarticletitle{Hybrid datafication of maintenance logs from AI-assisted human tags}. In \bibinfo{booktitle}{\emph{2017 IEEE International Conference on Big Data (Big Data)}}. \bibinfo{pages}{1769--1777}.
\newblock
\urldef\tempurl%
\url{https://doi.org/10.1109/BigData.2017.8258120}
\showDOI{\tempurl}


\bibitem[Shneiderman(2022)]%
        {shneiderman2022human}
\bibfield{author}{\bibinfo{person}{Ben Shneiderman}.} \bibinfo{year}{2022}\natexlab{}.
\newblock \bibinfo{booktitle}{\emph{Human-centered AI}}.
\newblock \bibinfo{publisher}{Oxford University Press}.
\newblock


\bibitem[Shortliffe et~al\mbox{.}(1975)]%
        {shortliffe_computer-based_1975}
\bibfield{author}{\bibinfo{person}{Edward~H. Shortliffe}, \bibinfo{person}{Randall Davis}, \bibinfo{person}{Stanton~G. Axline}, \bibinfo{person}{Bruce~G. Buchanan}, \bibinfo{person}{C.~Cordell Green}, {and} \bibinfo{person}{Stanley~N. Cohen}.} \bibinfo{year}{1975}\natexlab{}.
\newblock \showarticletitle{Computer-based consultations in clinical therapeutics: {Explanation} and rule acquisition capabilities of the {MYCIN} system}.
\newblock \bibinfo{journal}{\emph{Computers and Biomedical Research}} \bibinfo{volume}{8}, \bibinfo{number}{4} (\bibinfo{year}{1975}), \bibinfo{pages}{303--320}.
\newblock
\showISSN{0010-4809}
\urldef\tempurl%
\url{https://doi.org/10.1016/0010-4809(75)90009-9}
\showDOI{\tempurl}


\bibitem[Strickland(2019)]%
        {stricklandIBMWatsonHeal2019}
\bibfield{author}{\bibinfo{person}{Eliza Strickland}.} \bibinfo{year}{2019}\natexlab{}.
\newblock \showarticletitle{{{IBM Watson}}, Heal Thyself: {{How IBM}} Overpromised and Underdelivered on {{AI}} Health Care}.
\newblock  \bibinfo{volume}{56}, \bibinfo{number}{4} (\bibinfo{date}{04} \bibinfo{year}{2019}), \bibinfo{pages}{24--31}.
\newblock
\urldef\tempurl%
\url{https://doi.org/10.1109/MSPEC.2019.8678513}
\showDOI{\tempurl}


\bibitem[Tan et~al\mbox{.}(2006)]%
        {tanLiveCaptureReuse2006}
\bibfield{author}{\bibinfo{person}{Hai~Chen Tan}, \bibinfo{person}{Pat Carrillo}, \bibinfo{person}{Chimay Anumba}, \bibinfo{person}{John~M Kamara}, \bibinfo{person}{Dino Bouchlaghem}, {and} \bibinfo{person}{Chika Udeaja}.} \bibinfo{year}{2006}\natexlab{}.
\newblock \showarticletitle{Live Capture and Reuse of Project Knowledge in Construction Organisations}.
\newblock  \bibinfo{volume}{4}, \bibinfo{number}{2} (\bibinfo{date}{05} \bibinfo{year}{2006}), \bibinfo{pages}{149--161}.
\newblock
\showISSN{1477-8238}
\urldef\tempurl%
\url{https://doi.org/10.1057/palgrave.kmrp.8500097}
\showDOI{\tempurl}


\bibitem[Tao et~al\mbox{.}(2019)]%
        {tao_self-aware_2019}
\bibfield{author}{\bibinfo{person}{Wenjin Tao}, \bibinfo{person}{Ze-Hao Lai}, \bibinfo{person}{Ming~C. Leu}, \bibinfo{person}{Zhaozheng Yin}, {and} \bibinfo{person}{Ruwen Qin}.} \bibinfo{year}{2019}\natexlab{}.
\newblock \showarticletitle{A self-aware and active-guiding training \& assistant system for worker-centered intelligent manufacturing}.
\newblock \bibinfo{journal}{\emph{Manufacturing Letters}}  \bibinfo{volume}{21} (\bibinfo{date}{8} \bibinfo{year}{2019}), \bibinfo{pages}{45--49}.
\newblock
\showISSN{2213-8463}
\urldef\tempurl%
\url{https://doi.org/10.1016/j.mfglet.2019.08.003}
\showDOI{\tempurl}


\bibitem[Trappey et~al\mbox{.}(2022)]%
        {trappeyVRenabledEngineeringConsultation2022}
\bibfield{author}{\bibinfo{person}{Amy J.~C. Trappey}, \bibinfo{person}{Charles~V. Trappey}, \bibinfo{person}{Min-Hua Chao}, {and} \bibinfo{person}{Chun-Ting Wu}.} \bibinfo{year}{2022}\natexlab{}.
\newblock \showarticletitle{{{VR-enabled}} Engineering Consultation Chatbot for Integrated and Intelligent Manufacturing Services}.
\newblock   \bibinfo{volume}{26} (\bibinfo{date}{03} \bibinfo{year}{2022}), \bibinfo{pages}{100331}.
\newblock
\showISSN{2452-414X}
\urldef\tempurl%
\url{https://doi.org/10.1016/j.jii.2022.100331}
\showDOI{\tempurl}


\bibitem[Wagner(2006)]%
        {wagnerBreakingKnowledgeAcquisition2006c}
\bibfield{author}{\bibinfo{person}{Christian Wagner}.} \bibinfo{year}{2006}\natexlab{}.
\newblock \showarticletitle{Breaking the {{Knowledge Acquisition Bottleneck Through Conversational Knowledge Management}}}.
\newblock  \bibinfo{volume}{19}, \bibinfo{number}{1} (\bibinfo{date}{01} \bibinfo{year}{2006}), \bibinfo{pages}{70--83}.
\newblock
\showISSN{1040-1628}
\urldef\tempurl%
\url{https://doi.org/10.4018/irmj.2006010104}
\showDOI{\tempurl}


\bibitem[Wei et~al\mbox{.}(2022a)]%
        {wei2022emergent}
\bibfield{author}{\bibinfo{person}{Jason Wei}, \bibinfo{person}{Yi Tay}, \bibinfo{person}{Rishi Bommasani}, \bibinfo{person}{Colin Raffel}, \bibinfo{person}{Barret Zoph}, \bibinfo{person}{Sebastian Borgeaud}, \bibinfo{person}{Dani Yogatama}, \bibinfo{person}{Maarten Bosma}, \bibinfo{person}{Denny Zhou}, \bibinfo{person}{Donald Metzler}, \bibinfo{person}{Ed~H. Chi}, \bibinfo{person}{Tatsunori Hashimoto}, \bibinfo{person}{Oriol Vinyals}, \bibinfo{person}{Percy Liang}, \bibinfo{person}{Jeff Dean}, {and} \bibinfo{person}{William Fedus}.} \bibinfo{year}{2022}\natexlab{a}.
\newblock \bibinfo{booktitle}{\emph{Emergent Abilities of Large Language Models}}.
\newblock
\showeprint[arxiv]{2206.07682}~[cs.CL]


\bibitem[Wei et~al\mbox{.}(2022b)]%
        {weiChainofThoughtPromptingElicits2022}
\bibfield{author}{\bibinfo{person}{Jason Wei}, \bibinfo{person}{Xuezhi Wang}, \bibinfo{person}{Dale Schuurmans}, \bibinfo{person}{Maarten Bosma}, \bibinfo{person}{Brian Ichter}, \bibinfo{person}{Fei Xia}, \bibinfo{person}{Ed Chi}, \bibinfo{person}{Quoc~V. Le}, {and} \bibinfo{person}{Denny Zhou}.} \bibinfo{year}{2022}\natexlab{b}.
\newblock \showarticletitle{Chain-of-{{Thought Prompting Elicits Reasoning}} in {{Large Language Models}}}.
\newblock   \bibinfo{volume}{35} (\bibinfo{date}{12} \bibinfo{year}{2022}), \bibinfo{pages}{24824--24837}.
\newblock


\bibitem[Weinert et~al\mbox{.}(2022)]%
        {weinert2022designing}
\bibfield{author}{\bibinfo{person}{Tim Weinert}, \bibinfo{person}{Matthias Billert}, \bibinfo{person}{Marian~Thiel de Gafenco}, \bibinfo{person}{Andreas Janson}, {and} \bibinfo{person}{Jan~Marco Leimeister}.} \bibinfo{year}{2022}\natexlab{}.
\newblock \showarticletitle{Designing a Co-Creation System for the Development of Work-Process-Related Learning Material in Manufacturing}.
\newblock \bibinfo{journal}{\emph{Comput. Supported Coop. Work}} \bibinfo{volume}{32}, \bibinfo{number}{1} (\bibinfo{date}{1} \bibinfo{year}{2022}), \bibinfo{pages}{5–53}.
\newblock
\showISSN{0925-9724}
\urldef\tempurl%
\url{https://doi.org/10.1007/s10606-021-09420-5}
\showDOI{\tempurl}


\bibitem[Wellsandt et~al\mbox{.}(2022)]%
        {wellsandtHybridaugmentedIntelligencePredictive2022a}
\bibfield{author}{\bibinfo{person}{Stefan Wellsandt}, \bibinfo{person}{Konstantin Klein}, \bibinfo{person}{Karl Hribernik}, \bibinfo{person}{Marco Lewandowski}, \bibinfo{person}{Alexandros Bousdekis}, \bibinfo{person}{Gregoris Mentzas}, {and} \bibinfo{person}{Klaus-Dieter Thoben}.} \bibinfo{year}{2022}\natexlab{}.
\newblock \showarticletitle{Hybrid-Augmented Intelligence in Predictive Maintenance with Digital Intelligent Assistants}.
\newblock   \bibinfo{volume}{53} (\bibinfo{date}{01} \bibinfo{year}{2022}), \bibinfo{pages}{382--390}.
\newblock
\showISSN{1367-5788}
\urldef\tempurl%
\url{https://doi.org/10.1016/j.arcontrol.2022.04.001}
\showDOI{\tempurl}


\bibitem[Wieland et~al\mbox{.}(2016)]%
        {wielandRuleBasedManufacturingIntegration2016}
\bibfield{author}{\bibinfo{person}{Matthias Wieland}, \bibinfo{person}{Pascal Hirmer}, \bibinfo{person}{Frank Steimle}, \bibinfo{person}{Christoph Groeger}, \bibinfo{person}{Bernhard Mitschang}, \bibinfo{person}{Eike Rehder}, \bibinfo{person}{Dominik Lucke}, \bibinfo{person}{Omar~Abdul Rahman}, {and} \bibinfo{person}{Thomas Bauernhansl}.} \bibinfo{year}{2016}\natexlab{}.
\newblock \showarticletitle{Towards a {{Rule-Based Manufacturing Integration Assistant}}}. In \bibinfo{booktitle}{\emph{{{Factories of the Future in the Digital Environment}}}} (Amsterdam, 2016), \bibfield{editor}{\bibinfo{person}{E.~Westkamper} {and} \bibinfo{person}{T.~Bauernhansl}} (Eds.), Vol.~\bibinfo{volume}{57}. \bibinfo{publisher}{Elsevier Science Bv}, \bibinfo{pages}{213--218}.
\newblock
\showISSN{2212-8271}
\urldef\tempurl%
\url{https://doi.org/10.1016/j.procir.2016.11.037}
\showDOI{\tempurl}


\bibitem[Wurhofer et~al\mbox{.}(2018)]%
        {wurhoferReflectionsOperatorsMaintenance2018}
\bibfield{author}{\bibinfo{person}{Daniela Wurhofer}, \bibinfo{person}{Thomas Meneweger}, \bibinfo{person}{Verena Fuchsberger}, {and} \bibinfo{person}{Manfred Tscheligi}.} \bibinfo{year}{2018}\natexlab{}.
\newblock \showarticletitle{Reflections on {{Operators}}' and {{Maintenance Engineers}}' {{Experiences}} of {{Smart Factories}}}. In \bibinfo{booktitle}{\emph{Proceedings of the 2018 {{ACM International Conference}} on {{Supporting Group Work}}}} (New York, NY, USA) \emph{(\bibinfo{series}{{{GROUP}} '18})}. \bibinfo{publisher}{Association for Computing Machinery}, \bibinfo{pages}{284--296}.
\newblock
\showISBNx{978-1-4503-5562-9}
\urldef\tempurl%
\url{https://doi.org/10.1145/3148330.3148349}
\showDOI{\tempurl}


\bibitem[Xia et~al\mbox{.}(2023)]%
        {xia2023autonomous}
\bibfield{author}{\bibinfo{person}{Yuchen Xia}, \bibinfo{person}{Manthan Shenoy}, \bibinfo{person}{Nasser Jazdi}, {and} \bibinfo{person}{Michael Weyrich}.} \bibinfo{year}{2023}\natexlab{}.
\newblock \bibinfo{booktitle}{\emph{Towards autonomous system: flexible modular production system enhanced with large language model agents}}.
\newblock
\showeprint[arxiv]{2304.14721}~[cs.RO]


\bibitem[Xu et~al\mbox{.}(2021)]%
        {xuIndustryIndustryInception2021}
\bibfield{author}{\bibinfo{person}{Xun Xu}, \bibinfo{person}{Yuqian Lu}, \bibinfo{person}{Birgit Vogel-Heuser}, {and} \bibinfo{person}{Lihui Wang}.} \bibinfo{year}{2021}\natexlab{}.
\newblock \showarticletitle{Industry 4.0 and {{Industry}} 5.0—{{Inception}}, Conception and Perception}.
\newblock   \bibinfo{volume}{61} (\bibinfo{date}{10} \bibinfo{year}{2021}), \bibinfo{pages}{530--535}.
\newblock
\showISSN{0278-6125}
\urldef\tempurl%
\url{https://doi.org/10.1016/j.jmsy.2021.10.006}
\showDOI{\tempurl}


\bibitem[Yang et~al\mbox{.}(2019)]%
        {yangWhenKnowledgeNetwork2019b}
\bibfield{author}{\bibinfo{person}{Chi-Lan Yang}, \bibinfo{person}{Chien Wen~(Tina) Yuan}, {and} \bibinfo{person}{Hao-Chuan Wang}.} \bibinfo{year}{2019}\natexlab{}.
\newblock \showarticletitle{When {{Knowledge Network}} Is {{Social Network}}: {{Understanding Collaborative Knowledge Transfer}} in {{Workplace}}}.
\newblock   \bibinfo{volume}{3} (\bibinfo{date}{11} \bibinfo{year}{2019}), \bibinfo{pages}{164:1--164:23}.
\newblock
Issue CSCW.
\urldef\tempurl%
\url{https://doi.org/10.1145/3359266}
\showDOI{\tempurl}


\bibitem[Zhang et~al\mbox{.}(2021)]%
        {zhang2021an}
\bibfield{author}{\bibinfo{person}{Rui Zhang}, \bibinfo{person}{Nathan~J. McNeese}, \bibinfo{person}{Guo Freeman}, {and} \bibinfo{person}{Geoff Musick}.} \bibinfo{year}{2021}\natexlab{}.
\newblock \showarticletitle{"An Ideal Human": Expectations of AI Teammates in Human-AI Teaming}.
\newblock \bibinfo{journal}{\emph{Proc. ACM Hum.-Comput. Interact.}} \bibinfo{volume}{4}, \bibinfo{number}{CSCW3}, Article \bibinfo{articleno}{246} (\bibinfo{date}{1} \bibinfo{year}{2021}), \bibinfo{numpages}{25}~pages.
\newblock
\urldef\tempurl%
\url{https://doi.org/10.1145/3432945}
\showDOI{\tempurl}


\bibitem[Zhao et~al\mbox{.}(2023)]%
        {zhao2023survey}
\bibfield{author}{\bibinfo{person}{Wayne~Xin Zhao}, \bibinfo{person}{Kun Zhou}, \bibinfo{person}{Junyi Li}, \bibinfo{person}{Tianyi Tang}, \bibinfo{person}{Xiaolei Wang}, \bibinfo{person}{Yupeng Hou}, \bibinfo{person}{Yingqian Min}, \bibinfo{person}{Beichen Zhang}, \bibinfo{person}{Junjie Zhang}, \bibinfo{person}{Zican Dong}, \bibinfo{person}{Yifan Du}, \bibinfo{person}{Chen Yang}, \bibinfo{person}{Yushuo Chen}, \bibinfo{person}{Zhipeng Chen}, \bibinfo{person}{Jinhao Jiang}, \bibinfo{person}{Ruiyang Ren}, \bibinfo{person}{Yifan Li}, \bibinfo{person}{Xinyu Tang}, \bibinfo{person}{Zikang Liu}, \bibinfo{person}{Peiyu Liu}, \bibinfo{person}{Jian-Yun Nie}, {and} \bibinfo{person}{Ji-Rong Wen}.} \bibinfo{year}{2023}\natexlab{}.
\newblock \bibinfo{booktitle}{\emph{A Survey of Large Language Models}}.
\newblock
\showeprint[arxiv]{2303.18223}~[cs.CL]


\end{thebibliography}

\end{document}